\documentclass[pra,twocolumn,floatfix,superscriptaddress,longbibliography,notitlepage]{revtex4-2}
\usepackage{amssymb,amsmath,amsthm,color,graphicx,times,graphicx}
\usepackage{caption}
\usepackage{ragged2e}
\DeclareCaptionJustification{justified}{\justifying}
\usepackage{hyperref}
\usepackage{braket}
\usepackage{graphicx}
\usepackage{dcolumn}
\usepackage{appendix}
\usepackage{subcaption}
\usepackage{bm}
\DeclareMathOperator{\Tr}{Tr}

\providecommand{\openone}{\leavevmode\hbox{\small1\kern-4.3pt\normalsize1}}

 \usepackage{orcidlink}

\theoremstyle{plain}

\theoremstyle{definition}

\begin{document}
\title{Relativistic Quantum Thermometry in AdS Spacetime via Non-Markovian Temperature Sensing}

\author{Anass Hminat \orcidlink{0009-0007-3677-3952}}\affiliation{LPHE-Modeling and Simulation, Faculty of Sciences, Mohammed V University in Rabat, Rabat, Morocco.}
\author{Abdallah Slaoui \orcidlink{0000-0002-5284-3240}}\email{Corresponding author: abdallah.slaoui@um5s.net.ma}\affiliation{LPHE-Modeling and Simulation, Faculty of Sciences, Mohammed V University in Rabat, Rabat, Morocco.}\affiliation{Centre of Physics and Mathematics, CPM, Faculty of Sciences, Mohammed V University in Rabat, Rabat, Morocco.}
\author{Rachid Ahl Laamara \orcidlink{0000-0002-8254-9085}}\affiliation{LPHE-Modeling and Simulation, Faculty of Sciences, Mohammed V University in Rabat, Rabat, Morocco.}\affiliation{Centre of Physics and Mathematics, CPM, Faculty of Sciences, Mohammed V University in Rabat, Rabat, Morocco.}

\begin{abstract}
Quantum thermometry based on single-qubit sensor configurations enables the precise estimation of the temperature of a cosmological Anti-de Sitter (AdS) spacetime. In this work, we characterize the achievable estimation accuracy using the Quantum Fisher Information (QFI) and the associated quantum signal-to-noise ratio. For the first time, we introduce an ancillary Unruh-DeWitt detector between the sensor and the thermal bath, enhancing thermometric sensitivity by channeling temperature-dependent information into the probe qubit's coherence. We examine how detector acceleration in AdS space and the choice of boundary conditions modify the probe's thermal sensitivity. Despite the differing geometries, a unified phenomenology emerges: we characterize the scaling of the QFI with respect to temperature, detector energy gap, spacetime curvature, and interaction time. Finally, we identify optimal state preparation and measurement strategies that maximize the QFI, thereby establishing the fundamental limits of precision for non-Markovian sensing in curved spacetime.

\vspace{0.25cm}
\end{abstract}
\date{\today}

\maketitle

\section{Introduction}

The interplay between relativity and quantum information has given rise to an emerging and promising discipline, commonly referred to as relativistic quantum information \cite{1,Bouzaidi2025,Loulijat2026}. By incorporating relativistic effects, one can probe quantum phenomena under extreme spacetime conditions, particularly within curved backgrounds. Anti–de Sitter spacetime exhibits negative curvature and frequently serves as a theoretical setting in string theory, thereby providing crucial insights into quantum gravity and black hole physics. The Unruh–DeWitt (UDW) detector model \cite{2} furnishes a standard operational framework for examining the interaction between quantum fields and localized detectors in diverse spacetime geometries \cite{3}. One of the most notable predictions derived from this model is the Unruh effect: a detector uniformly accelerated through the Minkowski vacuum responds as if immersed in a thermal bath characterized by a Planckian spectrum, with an effective temperature proportional to the detector's proper acceleration \cite{4,5}. In quantum field theory, non-inertial observers need not agree on particle content: a field state that is vacuum for an inertial observer appears as a thermal ensemble to a uniformly accelerated (Rindler) observer—the well-known Unruh effect \cite{6,7}. Earlier studies have explored the quantum estimation of the Unruh temperature using a variety of accelerated probes \cite{8}. Relatedly, temperature estimation using a static two-level atom coupled to a massless scalar field in the presence of boundaries was addressed in \cite{9}, and the metrological performance of a uniformly moving two-level probe interacting with a massless scalar field has been analyzed in \cite{10,11,12}. That work found the estimation precision to be insensitive to the constant velocity itself, while decoherence induced by the field degrades accuracy over time.\par

To investigate quantum properties in AdS backgrounds, it is common to treat a Unruh–DeWitt detector linearly coupled to a massless scalar field as an open quantum system \cite{13}. In this paradigm, the detector constitutes the system, while the quantum field plays the role of the environment, inducing dissipation and decoherence. Concretely, the detector is modelled as a two-level quantum system with an energy gap $\Omega$ separating the states $|0\rangle_{D}$ and $|1\rangle_{D}$. Among the parameters of central relevance, temperature estimation plays a pivotal role in the characterization of complex environments \cite{41,42,43}. Determining the temperature of a quantum system is not only a fundamental problem but also one of considerable practical significance \cite{44,45}. Many quantum technologies operate optimally at ultralow temperatures in order to exploit fragile nonclassical features, thereby necessitating highly accurate temperature estimation schemes that introduce minimal back-action on the system \cite{46,47,48}. These requirements lie at the heart of quantum thermometry, a vibrant research field at the intersection of quantum metrology, quantum thermodynamics \cite{G1}, and the theory of open quantum systems \cite{49,50}.\par

Recent investigations into dynamical quantum thermometry, particularly those based on master-equation approaches, have shed light on both the equilibrium and nonequilibrium dynamics of quantum systems \cite{51,52}. Furthermore, collision models have been extensively employed to analyze thermalization mechanisms and temperature estimation protocols in open quantum systems \cite{53}. Complementarily, driving strategies have been proposed to enhance thermometric precision by coherently controlling energy spectra and transition processes within the probe system \cite{54,55}. Quantum probes \cite{56,57,58} have emerged as a powerful means of addressing this challenge. In particular, they provide a minimally invasive approach for estimating physical parameters of interest while inducing only negligible disturbance to the system under investigation \cite{61,62,63}, typically referred to as the quantum reservoir. The general protocol consists of preparing a simple quantum system—such as a single qubit or a pair of qubits—in a well-defined initial state. This probe is then allowed to interact with the target reservoir, during which information about the parameters of interest becomes encoded in its quantum state. Subsequent measurements performed on the probe enable the extraction of this information \cite{64,65,27,66}.\par

In this work, we report the first demonstration of probing an Unruh–DeWitt (UDW) detector that has been effectively insulated against dissipative dynamics and the thermal bath generated by AdS cosmological spacetime. This insulation is achieved by coupling the primary probe to an ancillary UDW detector: the ancillary system interacts directly with a scalar field and the thermal reservoir and, in doing so, conveys the reservoir’s thermal characteristics to the primary detector while mitigating direct dissipative influence on it. For the single-qubit configuration, we compare the quantum Fisher information (QFI) and the quantum signal-to-noise ratio (QSNR) for cases with and without an ancillary qubit. Specifically, the probe qubit exchanges only coherent coupling with an ancilla, while the ancilla itself undergoes dephasing due to contact with the sample; as a consequence, temperature information becomes imprinted in the probe’s off-diagonal elements. We show that increasing the probe–ancilla coupling enhances the achievable precision. The analysis is then extended to a pair of interacting qubits, where we examine both common and local bath scenarios and consider initial states that are either entangled or separable. When the composite system attains a steady state, the precision of temperature estimation is maximized. In the fully thermalized regime, the estimation efficiency—quantified via the QSNR—is found to be governed predominantly by the interqubit coupling strength. Thus, tuning this coupling affords practical control over thermometric performance, particularly at low temperatures.\par

The manuscript is organised as follows. In Section~\ref{X1}, we present the theoretical tools for a parameter-estimation protocol aimed at further characterising deviations from Markovian dynamics in the probe. we employ the information-backflow measure, which quantifies non-Markovianity through the temporary increase in state distinguishability. Section~\ref{X2}  provides a detailed description of the spacetime geometry under consideration. We then analyse the thermal behaviour emerging from the interaction between an Unruh–DeWitt (UDW) detector and a scalar field propagating in an \(\mathrm{AdS}_4\) spacetime. In Section~\ref{X3}, we examine the performance of a single-qubit detector, considered both independently and in conjunction with an auxiliary UDW system. Section~\ref{X4} investigates how the internal parameters of the spacetime influence coherence and information loss, which we quantify using the von Neumann entropy. In Section~\ref{X5}, we focus on the influence of these parameters on the thermal sensitivity of the system, as quantified by the quantum Fisher information (QFI) \(\mathcal{F}_T\). In particular, we identify the emergence of finite-time and finite-temperature maxima that define optimal operating conditions—denoted by \(t_{\rm opt}\) and \(T_{\rm opt}\)—at which the temperature estimation achieves its highest precision and the thermal sensitivity is maximised. In the section~\ref{X6}, we examine how non-Markovian dynamics, induced by the inter probe–ancilla coupling \(\kappa\), affect both the QFI, denoted \(\mathcal{F}_T\), and the QSNR, denoted \(R_T\). Our investigation encompasses both thermalised and non-equilibrium regimes, demonstrating how the coupling \(\kappa\) controls the thermal sensitivity of the auxiliary system and how this sensitivity is subsequently transferred to the probe. Additional technical details are provided in Appendix \ref{X7}, including an analyic solution of the master equation for an Unruh--DeWitt (UDW) detector in an $\mathrm{AdS}$ spacetime.

\section{Local quantum estimation theory}
\label{X1}
A parameter-estimation protocol aims to infer a quantity of interest (here denoted by \(T\)) indirectly, by processing the outcomes of a directly measurable observable \(X\). Let \(p(x\mid T)\) be the conditional probability density for obtaining outcome \(x\) when the true parameter value is \(T\). Given \(M\) independent measurement results \(x=\{x_1,x_2,\dots,x_M\}\), an estimator \(\hat{T}(x)\) maps the observed data to an estimate of the parameter ~\cite{a}. The average value of the estimator is
\begin{equation}
\overline{T} \;=\; \int dx\; p(x\mid T)\,\hat{T}(x),
\end{equation}
and the precision of the estimation procedure is quantified by the estimator variance
\begin{equation}
\mathrm{Var}\,\hat{T} \;=\; \int dx\; p(x\mid T)\,\bigl[\hat{T}(x)-\overline{T}\bigr]^{2}.
\end{equation}
Because the \(M\) measurements are performed on repeated, identically prepared copies, the joint outcome distribution factorizes as \(p(x\mid T)=\prod_{k=1}^{M} p(x_k\mid T)\).For any unbiased estimator (i.e. one for which \(\overline{T}\to T\) as \(M\to\infty\)), the variance is lower-bounded by the classical Cramér–Rao inequality ~\cite{C1,B4},
\begin{equation}
\mathrm{Var}\,\hat{T} \;\ge\; \frac{1}{M\,F(T)},
\end{equation}
where \(H(T)\) is the Fisher information associated with measurements of \(X\), defined by
\begin{equation}
H(T) \;=\; \int dx\; p(x\mid T)\,\bigl[\partial_{T}\log p(x\mid T)\bigr]^{2},
\end{equation}
and \(\partial_T\) denotes differentiation with respect to \(T\). The measurement that maximizes \(F(T)\) is the optimal one for estimating \(T\), while an estimator that attains the bound (3) is called efficient. The combination of the optimal measurement and an efficient estimator yields an optimal estimation scheme.Maximizing the Fisher information over all possible quantum measurements leads to the Quantum Fisher Information (QFI), denoted \(F(T)\) ~\cite{d,e,f}. The quantum Cramér–Rao bound then reads
\begin{equation}
\mathrm{Var}\,\hat{T} \;\ge\; \frac{1}{M\,F(T)}.
\end{equation}
The QFI can be expressed in terms of the spectral decomposition of the system density operator. Writing the state in diagonal form
\begin{equation}
\rho_{T} \;=\; \sum_{n} \rho_{n}\,\lvert\varphi_{n}\rangle\langle\varphi_{n}\rvert,
\end{equation}
the QFI is given by
\begin{equation}
F(T) \;=\; \sum_{p}\frac{(\partial_{T}\rho_{p})^{2}}{\rho_{p}}
\;+\; 2\sum_{n\ne m}\frac{(\rho_{n}-\rho_{m})^{2}}{\rho_{n}+\rho_{m}}\,
\bigl|\langle\varphi_{n}\!\mid\!\partial_{T}\varphi_{m}\rangle\bigr|^{2}.
\end{equation}
The first term in Eq.~(7) captures the dependence of the eigenvalues on the parameter and is often called the classical contribution to the QFI; the second term accounts for the parameter dependence of the eigenvectors and represents the genuinely quantum contribution. Note that the local quantum estimation framework used here assumes some prior, coarse knowledge of the parameter value \(T\) ~\cite{g}.

For a two-level probe it is convenient to use the Bloch representation. Any qubit state can be written as
\[
\hat{\rho} \;=\; \tfrac{1}{2}\big(\hat{I} + \mathbf{b}\!\cdot\!\hat{\boldsymbol{\sigma}}\big),
\]
where \(\hat{I}\) is the \(2\times2\) identity, \(\mathbf{b}=(b_x,b_y,b_z)\) is the real Bloch vector and \(\hat{\boldsymbol{\sigma}}=(\hat{\sigma}_x,\hat{\sigma}_y,\hat{\sigma}_z)\) are the Pauli matrices. In this representation the QFI for a (generally mixed) qubit state admits the closed form
\begin{equation}\label{qfi-bloch}
F(T)
\;=\;
\bigl\lVert\partial_{T}\mathbf{b}\bigr\rVert^{2}
\;+\;
\frac{\bigl(\mathbf{b}\!\cdot\!\partial_{T}\mathbf{b}\bigr)^{2}}{1-\lVert\mathbf{b}\rVert^{2}} \,.
\end{equation}

To evaluate the estimability of a parameter in a way that does not depend on its specific value, one introduces the signal-to-noise ratio (SNR)
\[
R_{T}\;=\;\frac{T^{2}}{\mathrm{Var}\,\hat{T}},
\]
which is larger for better estimators. Combining this definition with the quantum Cramér–Rao bound yields the quantum signal-to-noise ratio (QSNR)
\begin{equation}
R_{T} \;\le\; Q_{T} \equiv T^{2}\,H(T),
\end{equation}
so a larger \(Q_{T}\) indicates a parameter that can be estimated more effectively.To further characterise the non-Markovian behaviour of the probe we adopt an information-flow measure introduced by Breuer \emph{et al.}~\cite{aaa}. The degree of non-Markovianity is defined as
\begin{equation}
\mathcal{N}
\;=\;
\max_{\rho^{(P)}_{1,2}(0)}\;\int_{\sigma>0}\!dt\;\sigma(t)\,,
\end{equation}
where \(\sigma(t)=\dfrac{d}{dt}D(t)\) is the time derivative of the trace distance \(D(t)\) between two probe states. The trace distance is given by
\begin{equation}
D(t)=\tfrac{1}{2}\big\lVert \rho^{(P)}_{1}(t)-\rho^{(P)}_{2}(t)\big\rVert_{1}
=\tfrac{1}{2}\,\mathrm{Tr}\,\big\lvert \rho^{(P)}_{1}(t)-\rho^{(P)}_{2}(t)\big\rvert,
\end{equation}
with \(\lVert \cdot\rVert_{1}\) denoting the trace norm. Intuitively, \(D(t)\) quantifies the distinguishability of the two states: a monotonic decrease of \(D(t)\) corresponds to a continuous loss of information from the system to the environment (Markovian dynamics), whereas any temporal increase of \(D(t)\) signals a backflow of information from the environment into the system, which is the hallmark of non-Markovian dynamics ~\cite{bbb}. 

\noindent For a two-level probe it is customary (and sufficient) to optimise the measure over pairs of orthogonal pure states; a convenient choice are the eigenstates of the \(x\)-Pauli operator $|\pm\rangle$ ,which have been shown to maximise the Breuer–Laine–Piilo measure in many relevant cases~\cite{ccc}.
\section{Cosmological background }
\label{X2}
Reconciling quantum mechanics with general relativity remains among the principal open problems in contemporary physics. Recently, significant insights have emerged from the application of quantum information techniques to quantum field theory on both flat and curved backgrounds, a discipline often referred to as relativistic quantum information (RQI).  We will address relativistic quantum metrology in curved spacetime, restricting our attention to constant-curvature geometries, in particular to the Anti--de Sitter (AdS) backgrounds. Using Unruh--DeWitt detectors as local probes of the quantum vacuum, we build on earlier work that employed the Fisher information to estimate the expansion rate (i.e. the Hubble parameter) with an accelerated detector .In this section we shall discuss in detail the geometry of the spacetime under consideration and its parametrization, and we will thereafter examine the thermal behaviour inherent to the interaction between an Unruh–DeWitt (UDW) detector and a scalar field in an \(\mathrm{AdS}_4\) spacetime.

\subsection{Anti de-Sitter Space-time }
The four-dimensional anti–de Sitter (AdS$_4$) spacetimes may be realised as four-dimensional hyperboloids embedded in a five-dimensional ambient space  ~\cite{z1,z2}.  The line element of the embedding space is
\begin{equation}
\mathrm{d}s^{2} \;=\; -\,\mathrm{d}X_{0}^{2} + \mathrm{d}X_{1}^{2} + \mathrm{d}X_{2}^{2} + \mathrm{d}X_{3}^{2} - \mathrm{d}X_{4}^{2},
\end{equation}
with the upper sign corresponding to de Sitter space.  The (A)dS$_4$ manifold itself is obtained as the locus
\begin{equation}
-\,X_{0}^{2} + X_{1}^{2} + X_{2}^{2} + X_{3}^{2} -X_{4}^{2} \;=\; - \ell^{2},
\end{equation}
again taking the upper sign for dS$_4$.  The familiar static form of the (A)dS metric follows from these embeddings and can be written as
\begin{equation}
\mathrm{d}s^{2} \;=\; -\Big(1 - \frac{r^{2}}{\ell^{2}}\Big)\mathrm{d}t^{2}
+\Big(1 - \frac{r^{2}}{\ell^{2}}\Big)^{-1}\mathrm{d}r^{2} + r^{2}\mathrm{d}\Omega_{2}^{2},
\end{equation}
where $\ell$ denotes the characteristic (A)dS length scale  ~\cite{z3,z4}.  The static coordinates are obtained from the embedding coordinates by introducing the usual spherical parametrisation
\begin{equation}
X_{1} = r\sin\theta\sin\varphi,\qquad
X_{2} = r\sin\theta\cos\varphi,\qquad
X_{3} = r\cos\theta,
\end{equation} 
whereas for AdS$_4$ the corresponding relations read
\begin{equation}
X_{0} = \sqrt{\ell^{2}+r^{2}}\sin\!\big(\tfrac{t}{\ell}\big),
\qquad
X_{4} = \sqrt{\ell^{2}+r^{2}}\cos\!\big(\tfrac{t}{\ell}\big).
\end{equation}

The parameter $\ell=\sqrt{3/\Lambda}\equiv 1/k$ is the (A)dS length; in the de Sitter case the metric in static coordinates exhibits a coordinate singularity at the cosmological horizon $r=\ell$   ~\cite{z8}.

For de Sitter space it is often convenient to use comoving (flat) coordinates, in which the metric assumes the manifestly conformally flat form
\begin{equation}
\mathrm{d}s^{2} \;=\; -\mathrm{d}t^{2} + e^{2t/\ell}\big(\mathrm{d}x_{1}^{2}+\mathrm{d}x_{2}^{2}+\mathrm{d}x_{3}^{2}\big).
\end{equation}
These coordinates follow from the flat-space embedding via the transformations
\begin{align}
&T = \ell\sinh\!\big(\tfrac{t}{\ell}\big) + \frac{r^{2}}{2\ell}\,e^{t/\ell}, \hspace{1cm}
X_{0} = \ell\cosh\!\big(\tfrac{t}{\ell}\big) - \frac{r^{2}}{2\ell}\,e^{t/\ell}, \notag\\&
X_{1} = e^{t/\ell}x_{1}, \hspace{1cm} X_{2} = e^{t/\ell}x_{2}, \hspace{1cm} X_{3} = e^{t/\ell}x_{3},
\end{align}

\subsection{Unruh de Witt detector and thermal proprieties}

We employ the Unruh--DeWitt model of particle detectors \cite{100}, wherein the detector is represented by a two-level quantum system with energy eigenstates $\lvert 0 \rangle_{D}$ and $\lvert 1 \rangle_{D}$ separated by an energy gap $\Omega$ ~\cite{101}.  The detector follows a worldline $x(\tau)$ parametrised by its proper time $\tau$ and couples to a massless scalar field $\phi(x)$ through the interaction Hamiltonian (we adopt units and conventions as in the main text)
\begin{equation}
H_{I}(\tau) \;=\; \eta\;\big( e^{i\Omega\tau}\sigma_{+} + e^{-i\Omega\tau}\sigma_{-} \big)\otimes \phi\!\big[x(\tau)\big],
\end{equation}
where $\lambda$ denotes the coupling strength, and $\sigma_{+}= \lvert 1 \rangle_{D}\!\langle 0 \rvert_{D}$ and $\sigma_{-}= \lvert 0 \rangle_{D}\!\langle 1 \rvert_{D}$ are the detector ladder operators.  (In what follows we consider pointlike detectors without an explicit switching function)~\cite{102}

Consider next a real, massless scalar field $\varphi$ conformally coupled to curvature.  The action is chosen as
\begin{equation}
S \;=\; \int \mathrm{d}^{4}x\,\sqrt{-g}\,\left(\tfrac{1}{2}\,g^{\mu\nu}\nabla_{\mu}\varphi\nabla_{\nu}\varphi
- \tfrac{1}{12}\,R\,\varphi^{2}\right).
\end{equation}
Denote by $W(x,x')=\langle 0|\varphi(x)\varphi(x')|0\rangle$ the Wightman two-point function ~\cite{104,105}.  For a comoving observer in the conformal vacuum of dS$_4$ the Wightman function depends only upon the proper–time separation $\Delta\tau=\tau-\tau'$ and for the case $\eta=1$ we have :
\begin{equation}
W_{\mathrm{dS}}\!\big(x(\tau),x(\tau')\big)
\;=\; -\,\frac{1}{2\sqrt{2}\,\pi\ell}\,
\frac{1}{\sinh^{2}\!\big(\tfrac{\tau-\tau'}{\ell}-i\varepsilon\big)} ,
\end{equation}
so that the comoving trajectory is stationary.  The conformal vacuum respects the full symmetry of the de Sitter group.

The construction of quantum field theory on AdS$_4$ is more subtle because AdS is not globally hyperbolic and hence admits no Cauchy surface.  A standard procedure is to quantise on the Einstein static universe and then map the result to AdS via the conformal embedding ~\cite{106}.  One convenient representation of the Wightman function in AdS is
\begin{equation}
W_{\mathrm{AdS}}(x,x') \;=\; -\,\frac{1}{4\sqrt{2}\,\pi\ell}\,
\left[ \frac{1}{\sigma(x,x')} \;-\; \frac{\zeta}{\sigma(x,x')+2} \right],
\end{equation}
where $2\ell^{2}\sigma(x,x')$ denotes the squared geodesic distance between $x$ and $x'$ (computed from the embedding of the manifold) and where the parameter $\zeta\in\{1,0,-1\}$ selects Dirichlet, transparent, or Neumann boundary conditions, respectively ~\cite{107}.

To lowest nontrivial order in $\lambda$ the transition probability for the detector to make an excitation from $\lvert 0 \rangle_{D}$ to $\lvert 1 \rangle_{D}$ is proportional to the response function \cite{z9,z10}
\begin{equation}
\mathcal{G}(\omega) \;=\; \int_{-\infty}^{\infty}\mathrm{d}\tau \int_{-\infty}^{\infty}\mathrm{d}\tau'\;
e^{-i\omega(\tau-\tau')}\; W\!\big(x(\tau),x(\tau')\big),
\end{equation}
where $W(x,x')=\langle 0|\phi(x)\phi(x')|0\rangle$ is the Wightman two–point function of the field and $\omega$ is the detector energy gap (in this work $\omega=\Omega$).  

This system, commonly referred to as an Unruh--DeWitt (UDW) detector, follows a spacetime trajectory $x(\tau)$ parameterized by its proper time $\tau$, and is linearly coupled to a massless scalar quantum field $\chi(x)$.

For a special class of detector trajectories, known as stationary trajectories, the Wightman function depends solely on the proper time difference $\Delta \tau = \tau - \tau'$. In such cases, it is more convenient to introduce the detector’s response rate, defined as the transition probability per unit proper time , which is the Fourier transform of the Wightman function and captures the steady-state behavior of the detector--field interaction
\begin{equation}
\mathcal{G}(\omega) \;=\; \int_{-\infty}^{\infty}\mathrm{d}\Delta\tau\; e^{-i\omega\Delta\tau}\; W(\Delta\tau).
\end{equation}
The response per unit time is the principal quantity of interest when treating the UDW detector as an open quantum system and, in particular, when computing information–theoretic quantities such as the Fisher information derived from detector statistics. In anti--de Sitter space, a UDW detector perceives thermal radiation only when its proper acceleration exceeds a threshold value set by the AdS length scale \cite{z11}.

In this work we shall focus on three representative detector configurations: (i) a comoving detector in dS$_4$, (ii) a uniformly accelerated detector in dS$_4$, and (iii) a uniformly accelerated detector in AdS$_4$.

A uniformly accelerated detector in AdS$_4$ likewise perceives a nonvanishing temperature,
\begin{equation*}
T \;=\; \frac{\sqrt{a^{2}\ell^{2}-1}}{2\pi\ell},
\end{equation*}
and the corresponding response function may be compactly expressed as
\begin{equation}
\begin{split}
\mathcal{G}_{\mathrm{AdS}}^{\mathrm{accel}}(\omega)
= \frac{1}{e^{\omega/T}-1}
\Bigg[
&\frac{\omega}{2\pi}
- \frac{\zeta}{4\pi\ell \sqrt{4\pi^{2}T^{2}\ell^{2}+1}} \\
&\times \sin\!\left(
\frac{\omega}{\pi T}
\sinh^{-1}(2\pi T\ell)
\right)
\Bigg]\Theta(T).
\end{split}
\end{equation}
\begin{figure}
    \centering
\includegraphics[width=1.0\linewidth]{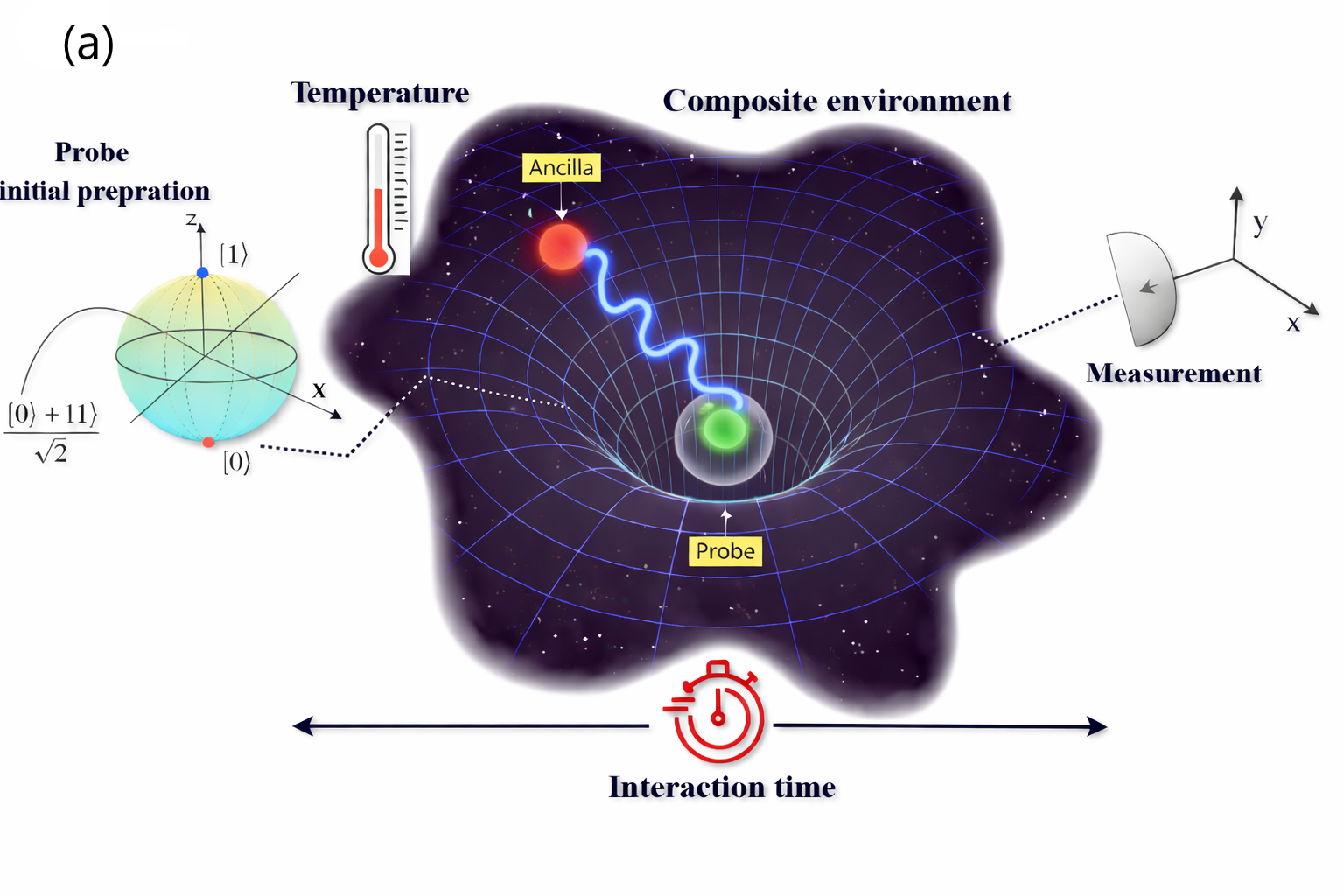}
\includegraphics[width=1.0\linewidth]{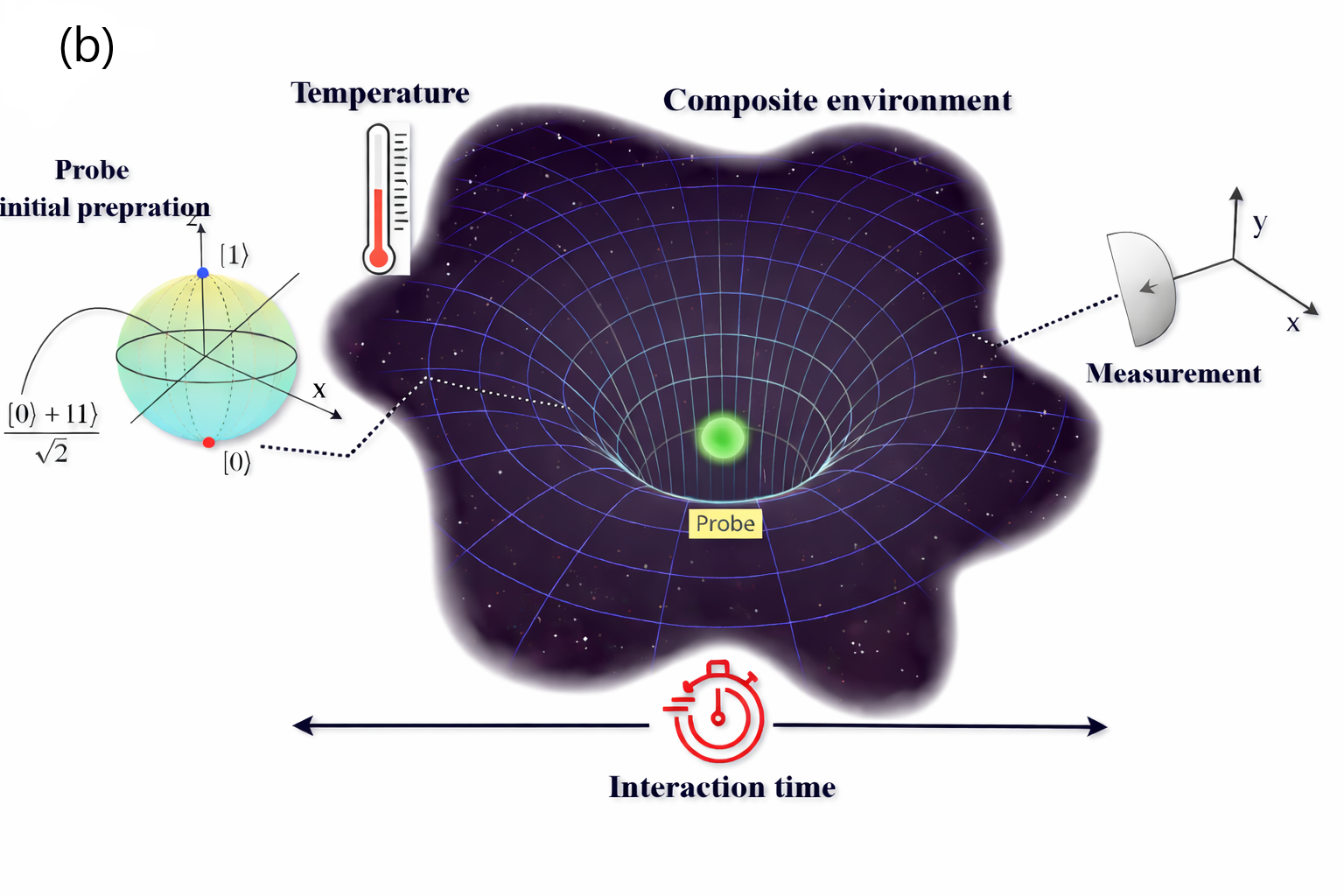}
\caption{Estimation theory is concerned with extracting an unknown parameter $T$ from a set of measurement outcomes while minimizing the associated estimation error. In the context of quantum thermometry, a quantum probe is employed to infer the temperature of an Ads space-time. The thermometric protocol considered here can be naturally divided into three fundamental stages. First, the probe is prepared in an appropriately chosen quantum state $\rho$. The probe is shielded by a protective bubble that prevents thermal exchanges with the spacetime background , and acquires temperature information through its interaction with an ancillary Unruh–DeWitt detector which effectively encodes the thermal characteristics of the cosmic environment. Finally, a measurement described by a set of positive operator-valued measure (POVM) elements is performed on the probe. The ultimate precision achievable in temperature estimation is fundamentally limited by the quantum Cramér--Rao bound.For the second protocol, the auxiliary particle and the protective bubble isolating the probe from thermal interactions with the environment are removed, allowing us to assess the impact of the auxiliary system on the precision of temperature estimation. }
    \label{fig00}
\end{figure}

where $\Theta(\cdot)$ denotes the Heaviside step function.  In particular, for the transparent boundary condition $\zeta=0$, Eq.~(21) reduces to the thermal response (19) appropriate to the comoving detector in dS spacetime.
A uniformly accelerated Unruh–DeWitt detector in anti–de Sitter space perceives a nonvanishing response for sufficiently large acceleration \cite{z12,z13}. The response function may be written most compactly in terms of the proper acceleration $a$ and the AdS length $\ell$. One convenient form is
\begin{equation}
\begin{split}
\mathcal{G}_{\mathrm{AdS}}(\omega)
=\Bigg[
&\frac{\omega}{2\pi}
-\frac{\zeta\,a}{4\pi}\,
\sin\!\Bigg(
\frac{2\omega\ell}{\sqrt{a^{2}\ell^{2}-1}}\,\sinh^{-1}\!\big(\sqrt{a^{2}\ell^{2}-1}\,\big)
\Bigg)
\Bigg] \\
&\times \Theta(a\ell-1)\;
\frac{1}{\exp\!\Big(\dfrac{2\pi\ell\,\omega}
{\sqrt{a^{2}\ell^{2}-1}}\Big)-1}\,.
\end{split}
\end{equation}

where $\Theta(\cdot)$ denotes the Heaviside step function and $\zeta\in\{-1,0,1\}$ selects the AdS boundary condition (Neumann, transparent, or Dirichlet respectively). The factor $\Theta(a\ell-1)$ encodes the existence of a threshold acceleration: the detector has a nonzero thermal response only when the proper acceleration exceeds the critical value
\begin{equation}
a_{\mathrm{crit}} \;=\; \frac{1}{\ell}\,,
\qquad\text{equivalently}\qquad
T \;=\; \frac{\sqrt{a^{2}\ell^{2}-1}}{2\pi\ell}\,,
\end{equation}
where the latter expression defines the effective temperature $T$ associated to the accelerated trajectory.

Equivalently, expressing the response directly in terms of the effective temperature $T$ yields the compact form
\begin{equation}
\begin{split}
\mathcal{G}_{\mathrm{AdS}}(\omega)
=\Bigg[
&\frac{\omega}{2\pi}
-\frac{\zeta}{4\pi\ell}\,
\sqrt{4\pi^{2}T^{2}\ell^{2}+1}\;
\sin\!\Bigg(
\frac{\omega}{\pi T}\,
\sinh^{-1}(2\pi T\ell)
\Bigg)
\Bigg] \\
&\times \Theta(T)\;
\frac{1}{e^{\omega/T}-1}\,.
\end{split}
\end{equation}

which is algebraically equivalent to Eq.~(2.25) upon using \(\;2\pi T\ell=\sqrt{a^{2}\ell^{2}-1}\).The first term in the square brackets of (2.25)–(2.27), namely $\omega/(2\pi)$, corresponds to the familiar thermal (Bose–Einstein) contribution. The denominator $e^{\omega/T}-1$ is the usual thermal factor. Unlike an accelerated detector in Minkowski spacetime (which experiences the Unruh temperature $T=a/2\pi$ for any nonzero $a$), an accelerated detector in AdS$_4$ is thermal only when the acceleration exceeds the critical value $a_{\mathrm{crit}}=1/\ell$. Below this threshold the thermal response vanishes. The second term inside the square brackets is a boundary–condition dependent correction to the purely thermal response. Its magnitude and sign depend on $\zeta$ and on the combination $a\ell$ (or, equivalently, on $T\ell$). In particular, for the transparent boundary condition $\zeta=0$ this correction vanishes and (2.27) reduces to the thermal response (2.19) appropriate to a comoving observer in dS$_4$.Physically, this behaviour reflects the fact that in AdS the spacetime curvature and the boundary at conformal infinity influence the detector response; only when the detector acceleration is large enough to overcome the AdS length scale does the usual thermal response appear.The equivalence between the AdS response with transparent boundary conditions and the comoving dS response (for the appropriate identification of parameters) will be exploited below when comparing Fisher information extracted by detectors in the two spacetimes.

\section{Temperature sensing via quantum probes}
\label{X3}
In this section we present our results on estimating the temperature of the thermal bath.Our goal is to employ UDW detector as quantum thermometers to estimate the equilibrium temperature T of the Ads cosmological space-time using various schemes. Specifically, we evaluate the performance of a single-qubit probe, studied both in isolation and assisted by an ancillary  UDW system in fig  \ref{fig00}.a-b.The analysis is performed by evaluating the time-evolution of the quantum Fisher information (QFI)  for fixed values of the UDW-qubit coupling strength $\kappa$. 
\subsection*{A. Single qubit with an ancillary assistant}
we first examine the evolution of an UDW ancilla  detector coupled to a scalar field that interact with an isolated qubit probe . The total Hamiltonian of the closed system may be written as
\begin{equation}
\label{eq:totalH}
H_T \;=\; H_{S} + H_{\chi} + H_{\rm int}\,,
\end{equation}
The Hamiltonian of the probe--ancilla subsystem $H_{S}$, is taken to be
\begin{equation}
H_{S} \;=\; \frac{\hbar}{2}\,\omega_{P}\,\sigma_{z}^{(P)}
\;+\; \frac{\hbar}{2}\,\omega_{A}\,\sigma_{z}^{(A)}
\;+\; \frac{\kappa}{2}\,\bigl(\sigma_{x}^{(P)}\sigma_{x}^{(A)}
+\sigma_{y}^{(P)}\sigma_{y}^{(A)}\bigr),
\label{eq:probe_ancilla_hamiltonian}
\end{equation}

where \(\omega_{P}\) (\(\omega_{A}\)) is the transition frequency of the probe (UDW ancilla), \(\kappa\) denotes the strength of the exchange-type coupling between the two-level systems, and \(\sigma_{x}^{(j)},\sigma_{y}^{(j)},\sigma_{z}^{(j)}\) are the Pauli matrices for subsystem \(j\in\{P,A\}\).
with  \(H_{\chi}\) is the free-field Hamiltonian for a massless scalar \(\chi(x)\) in the chosen curved background satisfying the covariant Klein--Gordon equation \(\Box\chi=0\), where \(\Box\equiv g^{\mu\nu}\nabla_{\mu}\nabla_{\nu}\) is the d'Alembertian associated with the chosen coordinate chart, and \(H_{\rm int}\) encodes the detector–field coupling , where the general expression for two Unruh--DeWitt (UDW) detectors interacting with a scalar field can be written in the following form:

\begin{equation}
\label{eq:Hint}
\begin{aligned}
H_{\mathrm{int}}^{\mathrm{UDW}}
=\, \eta \Big[
&\big(\sigma^{(P)}_{1}\!\otimes\!\mathbb{I}^{(A)}\big)\,
 \chi_{1}\!\left(t,x^{(P)}\right) \\
&+ \big(\mathbb{I}^{(P)}\!\otimes\!\sigma^{(A)}_{2}\big)\,
 \chi_{2}\!\left(t,x^{(A)}\right)
\Big] .
\end{aligned}
\end{equation}

 We initially couple the probe to an auxiliary particle and subsequently place the composite system in a cosmological background, which acts as a thermometer for the probe.
In the framework of our study, $\chi_{2}$ is set to zero since the probe is protected from the interaction with the scalar field and the spacetime background, similarly to the protocol presented in $X_{1}$.

In the weak-coupling (Born) regime, and under the usual Markov approximation, one can eliminate the field degrees of freedom and obtain a Markovian master equation for the detectors' reduced state. The resulting Gorini–Kossakowski–Sudarshan–Lindblad (GKSL, or Kossakowski–Lindblad) , with  \(x^{(m)}\) is the worldline position of detector \(m\).We denote by \(\rho_{\rm tot}(t)\) the density operator of the full system and by \(\rho_{PA}(t)=\mathrm{Tr}_{\chi}\big[\rho_{\rm tot}(t)\big]\) the reduced state of the two detectors (trace taken over the field).The master equation~(4) provides a powerful tool to analyze the long-time, asymptotic equilibrium states of the detectors. At large evolution times, these steady states emerge from a nontrivial interplay between dissipative effects induced by the environment in a curved spacetime (CST) background and the generation of quantum correlations driven by the Markovian dynamics of the detectors~\cite{27,65}.Assuming an initially factorised state \(\rho_{\rm tot}(0)=\rho_{PA}(0)\otimes\lvert 0\rangle\langle 0\rvert\), where \(\lvert 0\rangle\) is the field vacuum, the total state evolves under the von Neumann equation \(i\dot\rho_{\rm tot}=[H,\rho_{\rm tot}]\).In the weak-coupling, Born--Markov regime one may eliminate the field degrees of freedom and derive a Gorini--Kossakowski--Lindblad master equation for the detectors' reduced state:
\begin{equation}
\label{eq:master}
\frac{\partial\rho_{ab}(t)}{\partial t} \;=\; -i\big[H_{\rm eff},\rho_{ab}(t)\big] + \mathcal{L}\big[\rho_{ab}(t)\big]\,.
\end{equation}
The effective Hamiltonian \(H_{\rm eff}\) contains Lamb-shift corrections and can be written as
\begin{equation}
\label{eq:Heff}
H_{\rm eff} \;=\; H_{S} - \frac{i}{2}\sum_{m,n=1}^{2}\sum_{i,j=1}^{3} \Lambda^{(mn)}_{ij}\,\sigma^{(m)}_{i}\sigma^{(n)}_{j}\,,
\end{equation}
Here, $\tilde{H}_{\mathrm{eff}}$ denotes the effective Hamiltonian of the system, where
$\Omega$ represents the renormalized energy-level splitting of the qubit. The imaginary
term  corresponds to the Lamb shift arising from the
interaction with the environment. In our regime, this contribution can be
safely neglected, as it is much smaller than $H_T$. While the dissipator assumes the Kossakowski form
\begin{equation}
\label{eq:dissipator}
\mathcal{L}[\rho] \;=\; \frac{1}{2}\sum_{m,n=1}^{2}\sum_{i,j=1}^{3}\Lambda^{(mn)}_{ij}
\Big(2\,\sigma^{(n)}_{j}\rho\,\sigma^{(m)}_{i} - \{\sigma^{(m)}_{i}\sigma^{(n)}_{j},\rho\}\Big),
\end{equation}
where the real coefficients \(\Lambda^{(mn)}_{ij}\) are elements of the Kossakowski matrices determined by field correlation functions.

At leading order in perturbation theory, the probability for the detector to undergo a transition from its ground state $\ket{0}_D$ to the excited state $\ket{1}_D$ is governed by the so-called response function,

Explicitly, the relevant field correlations are the Wightman functions
\begin{equation}
\label{eq:Wightman}
\mathcal{G}^{(mn)}(t-t') \;=\; \langle 0|\chi\big(t,x^{(m)}\big)\,\chi\big(t',x^{(n)}\big)|0\rangle,
\end{equation}
whose Fourier transforms enter the dissipative coefficients:
\begin{equation}
\label{eq:Fourier}
\mathcal{G}^{(mn)}(\omega) \;=\; \int_{-\infty}^{\infty} \!dt\,e^{i\omega t}\,\mathcal{G}^{(mn)}(t).
\end{equation}

In the regime of sufficiently small interatomic separation one may set all Kossakowski matrices equal, \(\Lambda^{(aa)}_{ij}=\Lambda^{(bb)}_{ij}=\Lambda^{(ab)}_{ij}=\Lambda^{(ba)}_{ij}\equiv \Lambda_{ij}\). A convenient parametrisation is
\begin{equation}
\label{eq:KossParam}
C_{ij} \;=\; A\,\delta_{ij} - i\,B\,\varepsilon_{ij3} + C\,\delta_{i3}\delta_{j3},
\end{equation}
with
\begin{align}
A &= \tfrac{\eta^2}{2}[\mathcal{G}(\omega)+\mathcal{G}(-\omega)] \\
B &= \tfrac{\eta^2}{2}[(\mathcal{G}(\omega)-\mathcal{G}(-\omega)] \\
C  &= \eta^2[\mathcal{G}(0)-A]
\end{align}

Solving the master equation \eqref{eq:master} then yields the long-time, asymptotic state of the two detectors, which may be expressed in Bloch form and whose properties reflect the competition between environmental dissipation in the chosen curved-space background and the coherence and correlations generated dynamically by the Markovian evolution.The dynamical density operator of the probe (by tracing the degrees of freedom of the UDW auxiliary system and the scalar field) have the following form :
\[
\rho_{P}(t)=\Tr_{\chi A}\!\big[\rho_{\rm tot}(t)\big]
\]
 As the combined probe + UDW auxiliary system + scalar field is closed, its time evolution is governed by the von Neumann equation . Here $\mathcal{L}[\rho_p]=0$ since the probe is deliberately decoupled from the
cosmological thermal bath and therefore experiences no direct dissipative
dynamics. Temperature information is instead acquired indirectly via an
ancillary Unruh--DeWitt detector, which interacts with the field and thus
functions as a localized thermometer. This arrangement endows the probe with
thermal sensitivity while shielding it from decoherence and potential
information loss associated with direct coupling to the AdS spacetime.

\section{Quantum Thermometry Protocols and coherence protection }
\label{X4}
All realistic quantum systems are inevitably coupled to their surroundings, and these system-environment interactions typically induce loss of coherence and degradation of accessible information .  Paradoxically,information loss can be significantly reduced by monitoring the open-system dynamics into a sensitive probe for estimating environmental parameters , prior to developing the technical apparatus of quantum thermometry. In the previous section we described the thermal properties of the AdS background and illustrated the protocols employed in Fig.~\ref{fig00}.We will in this section  compare the two protocols to determine which is the most optimized and effective for temperature estimation, while selecting the optimal preparation state for the auxiliary Unruh–DeWitt (UDW) detector. We then investigate how the spacetime’s internal parameters affect coherence and the loss of information, which we quantify using the von Neumann entropy.

\begin{figure}
    \centering
\includegraphics[width=1.0\linewidth]{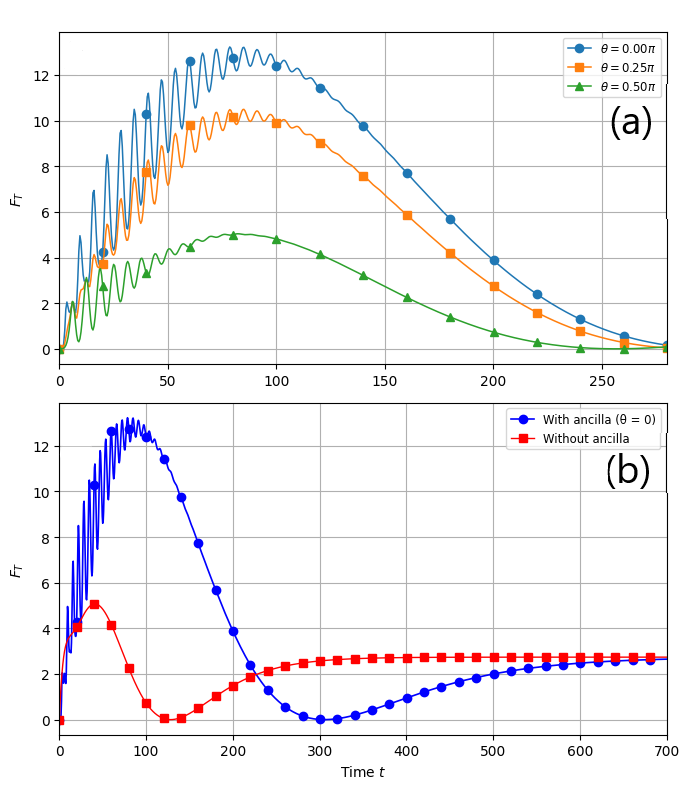} 
    \caption{We perform a comparative study of the optimal preparation state \(\theta\) of the auxiliary Unruh–DeWitt (UDW) system and assess its impact on thermometric efficiency, comparing it with the case where no auxiliary system is present.Top panel: Quantum Fisher information (QFI) $F_T$ computed from the reduced state of the probe qubit , for several initial preparations of the ancilla. Parameters: $T=0.4$, $\zeta=1$,$\kappa=0.25$ and $\eta=1$.
Bottom panel: QFI of the probe in the absence of the ancilla (red, solid) and in the presence of the ancilla (blue, dashed) for  $T=0.4
$, $\zeta=1$ and $\eta=1$. 
All frequencies and energies are expressed in units of $\omega_A=\omega_P=1$. }
    \label{fig0}
\end{figure}

\subsection{Quantum thermometry of a single-Unruh de Witt detector }

When the environment takes the form of a complex quantum reservoir with many degrees of freedom, accurately inferring its properties becomes a nontrivial task. Quantum probes offer a practical and minimally invasive route to this problem: a simple quantum system is prepared in a known initial state, allowed to interact with the reservoir so that relevant information is encoded in the probe's state, and subsequently measured to recover the target parameters . Temperature estimation is a particularly important example of this approach, since determining the reservoir temperature is both conceptually fundamental and of direct practical relevance for characterizing complex environments . We consider the case of a probe qubit initially prepared in a superposition state and directly introduced into the AdS spacetime $\ket{+} = \frac{1}{\sqrt{2}}\big(\ket{0} + \ket{1}\big)$.

\begin{figure}
    \centering
\includegraphics[width=1.0\linewidth]{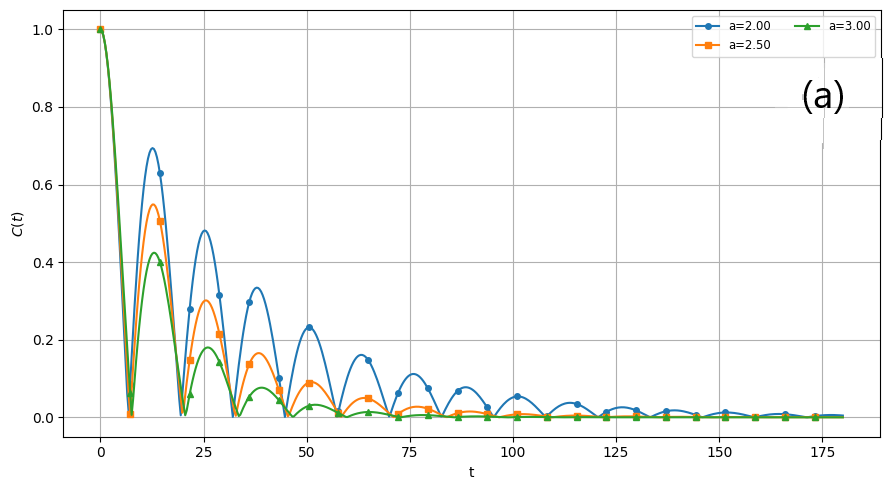} 
\includegraphics[width=1.0\linewidth]{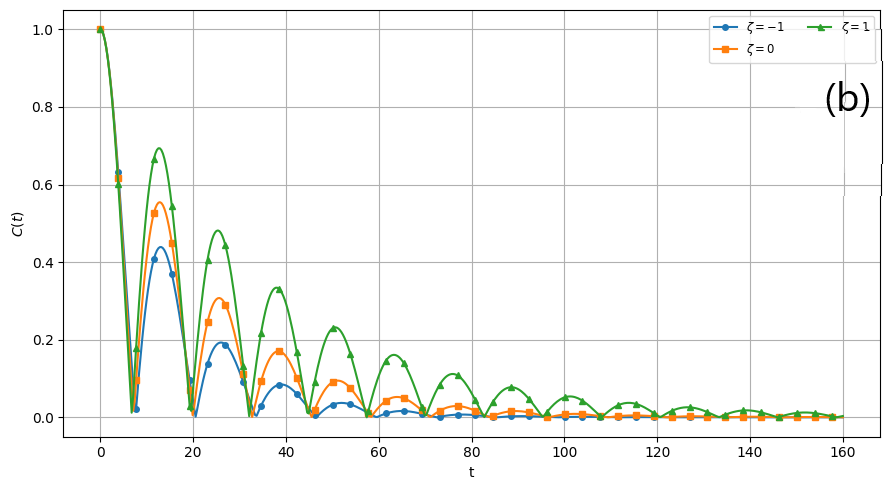} 
\includegraphics[width=1.0\linewidth]{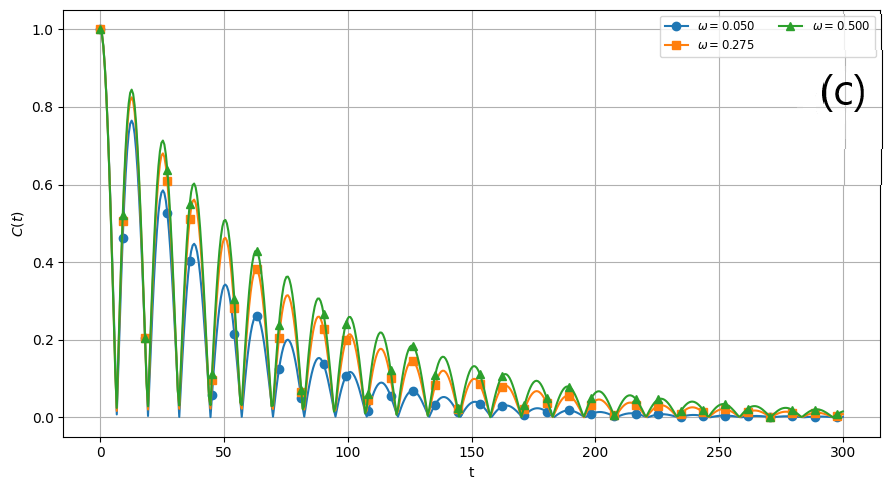}
    \caption{Top panel: Effects of acceleration and boundary conditions on quantum coherence. 
Left panels: Time evolution of the quantum coherence \(C_{\mathrm{AdS}}\) in AdS spacetime for different values of the  temperature \(T\), detector energy gap \(\omega\), and initial-state parameter \(\theta\)=0. Distinct coloured curves correspond to different detector accelerations: the blue solid line denotes the inertial case \(a=2\), the orange line corresponds to \(a=2.5\), and the green line represents \(a=3\). 
medium panels: Quantum coherence \(C_{\mathrm{AdS}}\) in AdS spacetime under varying temperature \(T\), energy gap \(\omega\), curvature \(k\). The coloured curves indicate different boundary conditions: Dirichlet (\(\zeta=1\), green solid line), transparent (\(\zeta=0\), orange line), and Neumann (\(\zeta=-1\), blue line). In several panels, the three curves coincide, resulting in a single visible trajectory due to complete overlap.In the lower panel, we plot the coherence $C_{\mathrm{Ads}}$ as a function of time $t$ for different energy gaps $\omega$. The green curve corresponds to $\omega = 0.5$, the orange curve to $\omega = 0.275$, and the blue curve to $\omega = 0.05$. }
    \label{fig1}
\end{figure}

In contrast, when the probe qubit interacts with the bath indirectly through the UDW ancillary systemv (in fig  \ref{fig00}-a), as in the single-qubit scenario without an auxiliary system (in fig  \ref{fig00}-b), we prepare the probe in the same initial state to provide a reference for comparison.We assume the probe to be initially prepared in its superposed state $\ket{+}_P$~\cite{44}. The ancillary system is taken to be initially in a pure state, parametrized by an angle $\theta$, which allows us to optimize the information transfer between the probe and the environment.

\begin{equation}
\label{eq:ancilla_init_eng}
\lvert \psi_A(0)\rangle = \cos\!\left(\frac{\theta}{2}\right)\lvert 0_A\rangle
+ \sin\!\left(\frac{\theta}{2}\right)\lvert 1_A\rangle.
\end{equation}

To determine the ancilla initial preparation that maximizes the probe sensitivity, we compute the QFI of the probe's reduced state as a function of the interaction time $t$ for several values of $\theta$ . Our results indicate that, in the indirect-coupling scenario, the optimal choice is an UDW ancilla in the ground state $\lvert 0_A\rangle$  with an probe prepared in the balanced superposition.Since measurements are local on the probe (which is protected by a bubble), the probe's reduced density matrix after tracing out the scalar field and the ancilla reads
\begin{equation}
\label{eq:rhoP_eng}
\rho_P(t)=\mathrm{Tr}_A\{\rho_{PA}(t)\}
=\dfrac{1}{2}
\begin{pmatrix}
1+\mathbb{A}(t) & 2\mathbb{B}(t)^* \\[6pt]
2\mathbb{B}(t) & 1-\mathbb{A}(t)

\end{pmatrix},
\end{equation}
where $\mathbb{A}(t)$ denotes the population imbalance and $\mathbb{B}(t)$ the coherence term. These quantities depend on the model parameters: the coupling strength $\kappa$, the bath parameter $\eta$, and the temperature $T$. The figure \ref{fig0} displays the quantum Fisher information (QFI) for two setups: (i) a probe directly coupled to the thermal bath  (solid red curve), and (ii) an indirect configuration in which the probe interacts only with an ancilla  (blue red curve) that is itself coupled to the sample The lower panel of  . In fig\ref{fig0}-a the QFI is plotted as a function of the interaction time \(t\) , we see that at very short times the directly coupled probe carries the bulk of the thermometric information and therefore exhibits the larger QFI. As the interaction time grows, however, the indirect (probe–ancilla) scheme accumulates information and eventually matches — and for long times surpasses — the QFI of the direct-coupling case, yielding superior temperature sensitivity.
The ancilla-mediated architecture thus allows information to be stored and funneled into the probe over much longer durations, leading to a markedly higher asymptotic thermometric precision. Importantly, after an initial transient the vast majority of this information can be retrieved by a local measurement on the probe alone, which considerably simplifies the readout procedure.
The coherence and population components of the unruh de-witt probe  encode essential information about the temperature of the sample. 
Through an indirect interaction mediated by an ancilla, the quantum state of the probe is modified, leading its coherence and population dynamics to carry signatures of the sample’s thermal properties. 
Consequently, monitoring these quantum characteristics enables the extraction of temperature information within this indirect sensing scheme.

Notably, the presence of an intermediate system allows the probe qubit to accumulate information over extended time scales, leading to a substantial improvement in thermometric sensitivity. 
A remarkable aspect of this approach is that, beyond a certain initial evolution time, nearly all the information encoded in the quantum state can be retrieved through local measurements performed solely on the probe qubit.

In both scenarios, the amount of extracted information is essentially the same and converges to an identical asymptotic value at long interaction times. However, the convergence occurs faster for the probe alone, since it is not affected by the coupling to the ancilla. The presence of the ancilla slightly slows down the information flow, leading to a delayed approach to the steady value.

\subsection{Coherence uncertainty }

In the context of temperature estimation of a thermal reservoir, an increase in quantum uncertainty reflects a growing incompatibility between the system’s state and the relevant energy observable. This incompatibility reduces the sensitivity of the state to temperature variations and, consequently, leads to a decrease in the Quantum sensing of the probe
We will nonetheless discuss the results concerning quantum coherence, while referring to \cite{a}, for a detailed analysis of the quantum uncertainty. where these aspects have been thoroughly investigated in de~Sitter and anti--de~Sitter spacetimes. In particular, those studies analyze the measured uncertainty and quantum coherence of an Unruh--DeWitt detector in de~Sitter spacetime, focusing on the influence of the temperature $T$, the energy gap $\Omega$, and the acceleration $a$ on the dynamical behavior of uncertainty and coherence for both comoving and accelerated detectors.

\begin{figure}
    \centering
\includegraphics[width=1.0\linewidth]{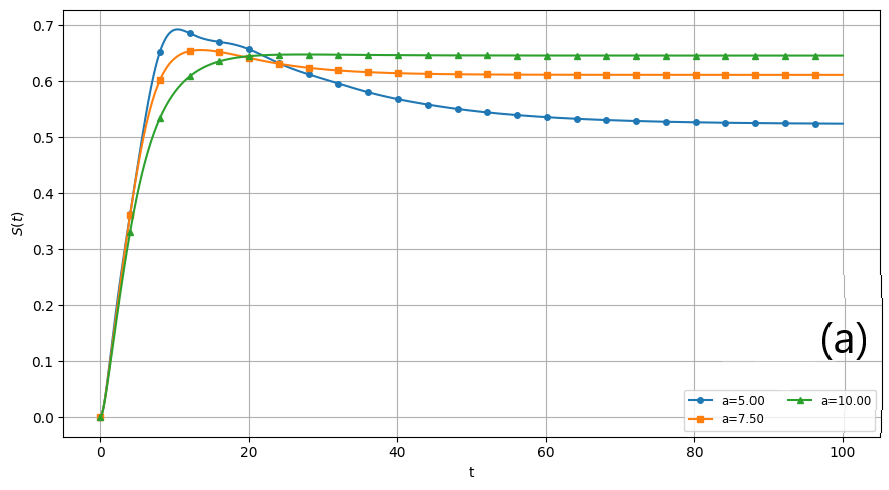} 
\includegraphics[width=1.0\linewidth]{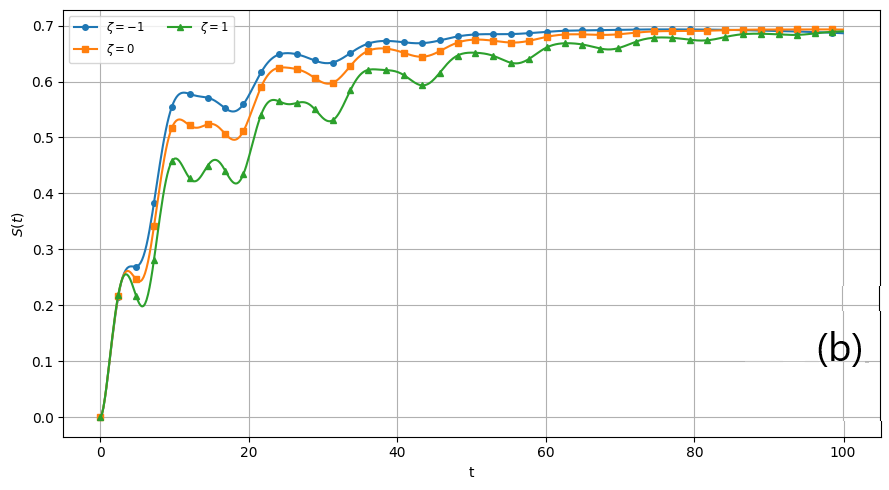} 
\includegraphics[width=1.0\linewidth]{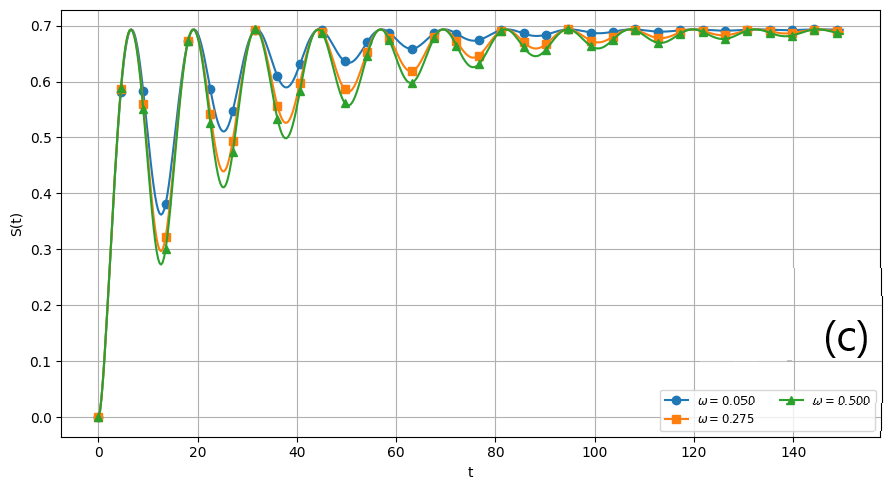}
    \caption{Top panel: The von Neumann entropy $S(t)$ as functions of the interaction time $t$ for different energy gaps $\omega$ ,  accelerations \(a\) and boundary conditions \(\zeta\) on the probe. 
Left panels: Time evolution of the Von Neumann entropy \(S_{\mathrm{AdS}}\) in AdS spacetime for different values of the  temperature \(T\), detector energy gap \(\omega\), and initial-state parameter \(\theta\)=0. Distinct coloured curves correspond to different detector accelerations: the green solid line denotes the inertial case \(a=10\), the orange line corresponds to \(a=7\), and the blue line represents \(a=5\).in the medium panel the coloured curves indicate different boundary conditions: Dirichlet (\(\zeta=1\), green solid line), transparent (\(\zeta=0\), orange), and Neumann (\(\zeta=-1\), blue line). In several panels, the three curves coincide, resulting in a single visible trajectory due to complete overlap.In the lower panel, we plot  $S_{\mathrm{Ads}}$ as a function of time $t$ for different energy gaps $\omega$. The green curve corresponds to $\omega = 0.5$, the orange curve to $\omega = 0.275$, and the blue curve to $\omega = 0.05$. }
    \label{fig2}
\end{figure}

\begin{figure*}[t] 
        \includegraphics[width=18cm]{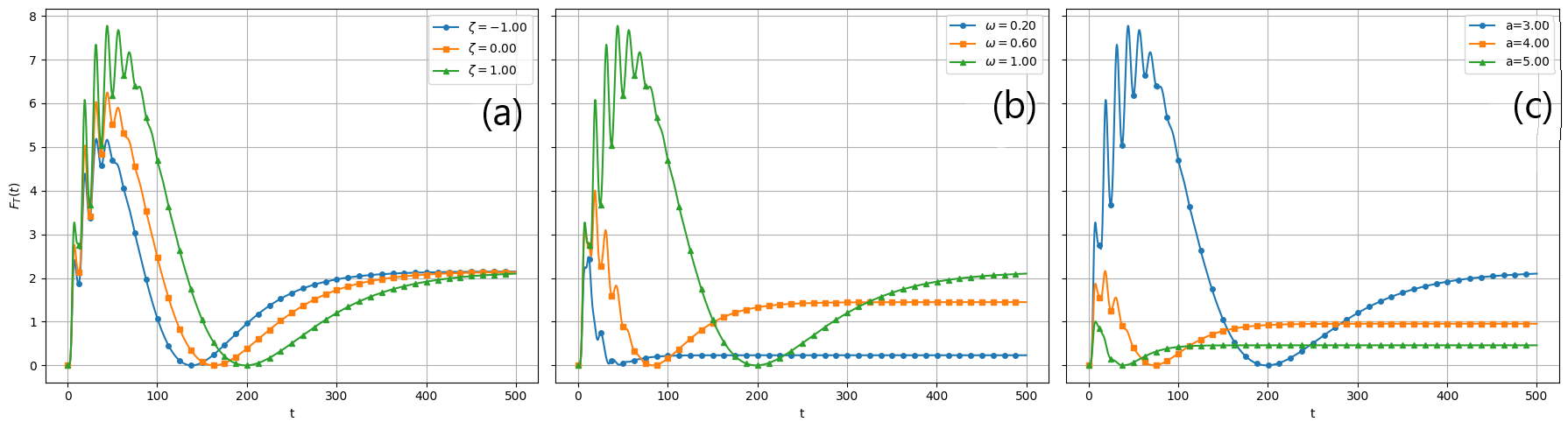} 
    \caption{A gallery of representative time evolutions of the QFI AdS$_4$ for different boundary condition (\(\zeta=1\)) , energy gap \(\omega\) and acceleration a. }
    \label{fig3}
\end{figure*}

The  quantum coherence detected by the  Unruh--DeWitt probe, \(C_{AdS}(\tau)\) in fig  \ref{fig1}.a-c, relaxes to a stationary nulle value \(C_{s}\) as \(\tau\to\infty\). Physically, \(C_{dS}\) measures the total off-diagonal weight remaining in the probe after interaction with the field; it thus governs the visibility of interference effects and the probe's ability to store and transmit genuinely quantum resources .It is clear that the thermalization reduces coherence If the detector experiences acceleration, a natural question arises: how does acceleration affect the uncertainty of the estimation process and the coherence of the quantum system? Motivated by this issue, we now turn to the analysis of accelerated detectors in fig  \ref{fig1}-a. In particular, we investigate how acceleration modifies the dynamical behavior of the detectors and influences their coherence properties and estimation performance . A detector undergoing uniform acceleration in anti–de Sitter space also responds thermally under suitable conditions. A salient feature of the AdS result is the presence of a critical acceleration: the detector exhibits a nonzero thermal response only when \(a>1/\ell\), i.e. only for accelerations exceeding the inverse AdS radius. This contrasts with the flat-space Unruh effect, where any nonzero acceleration yields a thermal response \(T=a/(2\pi)\). By increasing the acceleration  the temperature \(T\) rises and lowers the coherence \(C_{s}\) and accelerates the decay of \(C_{AdS}(\tau)\), whereas Fig.~\ref{fig1}-b shows that increasing the detector gap \(\omega\) tends to protect coherence and to reduce the long-time uncertainty , the boundary conditions that shape the geometry of our spacetime also influence quantum coherence, since coherence is sensitive to the parameter \(\zeta\) 
, where larger \(\zeta\) enhances the residual coherence and reduces the long-time uncertainty as demonstrated in  fig  \ref{fig1}-c. By contrast, at high temperature the influence of the boundary conditions is suppressed because the boundary-related contribution  tends to zero . Consequently the curves for different \(\zeta\) become indistinguishable.Physically, increasing \(T\) amplifies thermal noise and excitations, which quickly wash out boundary-induced distinctions.Hence the eventual uncertainty and coherence depend only on the ratio \(\omega/T\): increasing \(\omega/T\) yields lower stationary uncertainty and higher stationary coherence. This observation explains the monotonic dependence and the approximate anti-correlation between residual quantumness and measurement uncertainty in \cite{ref108}.

To investigate the loss of information, we examine the correlation and the transfer of information between the system and its surrounding environment and its competition with coherence in the dynamics of the probe. In particular, we focus on the entanglement entropy, which serves as a fundamental measure of quantum correlations and the information content within the system in fig \ref{fig2}.a-c. For the present setup, this quantity specifically characterizes the entanglement between the qubit and its environment.

In our setup, relaxes to a stationary value \(S_{s}\) that depends solely on the ratio \(\omega/T\) similarly to the coherence. At early times \(S(t)\) can display oscillations for probes prepared in superposition states, owing to an oscillatory contribution , these oscillations are amplified by larger energy gaps \(\omega\) . Increasing the acceleration \(a\) suppresses transient oscillations  and raises the asymptotic uncertainty, whereas nonzero detector acceleration only affects the short-time behaviour and does not modify the long-time value \(S(t)\).
Notably, larger values of the boundary parameter \(\zeta\) typically enhance the steady-state coherence and lower the steady-state uncertainty, whereas the effect of the boundary conditions is only observable at finite times and vanishes completely upon thermalization, resulting in a similar behavior for the three curves that becomes independent of $\zeta$.Consequently, \(S(t)\) serves as a robust indicator of the competition between coherent dynamics and thermal decoherence in curved-spacetime probes.

These results suggest a reduction of system--environment entanglement by seeking to optimize dissipation of coherence and information through a suitable choice of the internal parameters. The ADS-geometry appears to provide moderate protection via a gain–loss balance, whereas \(\zeta=1\), weak accelerations, and a larger energy gap \(\omega\) offer the strongest shielding—possibly because thermalization and information loss reduce the effective system–environment coupling.

\section{Thermal dynamics of the probe and optimal performance}
\label{X5}
In the previous section we examined how the internal parameters of our cosmological reservoir influence quantum coherence and information dissipation, originating from the residual entanglement between the auxiliary Unruh--DeWitt (UDW) detector and the AdS cosmological horizon. In this section we analyze the impact of those parameters on the system's thermal sensitivity, as quantified by the Quantum Fisher Information (QFI), $\mathcal{F}_T$. We then identify finite-time and finite-temperature peaks that establish optimal operating conditions — denoted $t_{\rm opt}$ and $T_{\rm opt}$ — where the temperature estimate attains its highest precision and the thermal sensitivity is maximal.which will be examined in greater detail in the next section.
\subsection{Metrological sensing with auxiliary Unruh-DeWitt detectors}

As in the preceding analysis, we first focus on the qualitative features of the Fisher
information $F(\tau)$ in Anti--de Sitter spacetime across the three
admissible boundary conditions in AdS , notably, it coincide with the case  found in the de Sitter case at ($\zeta = 0$)  , by contrast the introduction of nontrivial boundary conditions ($\zeta = \pm 1$) enriches
the phenomenology by activating additional degrees of freedom, thereby giving rise to new
dynamical regimes.  A key implication of this analysis is that the
long-time (asymptotic) Fisher information depends solely on the temperature $T$ and the
detector energy gap $\omega$, for all observers, and remains insensitive to the specific details of their trajectories.We will further also investigate how the energy gap $\omega$ and the boundary-condition factor $\zeta$ influence the thermal estimation sensed by the probe.

The QFI in fig \ref{fig3}-a  increases monotonically with larger values of \(\zeta\). Thus, there exists a clear trade-off controlled by the cosmological boundary \(\zeta\) , and  achieving a higher optimal QFI requires a longer optimal encoding time.

The relaxation toward constant stationary values as \(\tau\) grows, reflecting thermalisation of the Unruh–DeWitt detector. At low temperatures the short-time differences between curves for different \(\zeta\) can be significant: 
larger \(\zeta\) tends to enhance the residual coherence and reduce the uncertainty during the transient dynamics. However,upon thermalization, the system reaches a static equilibrium state that is independent of the spacetime boundaries. Consequently, the temperature estimates corresponding to the three values of $\zeta$ become indistinguishable and asymptotically converge to a single value.

We now turn to the study of the QFI in fig \ref{fig3}.b-c with respect to the acceleration $a$ and the energy gap $\omega$, and examine their effects on the probe.

An increase in acceleration enhances the temperature perceived by the probe particle.Which in turn hampers the estimation process and degrades the thermal sensitivity , where the information transfer between the auxiliary system and the probe is significantly perturbed . Physically, this behaviour reflects the dominance of thermal noise and excitation at large accelerations: the system rapidly approaches thermal equilibrium and the influence of the detector's acceleration \(a\) is confined to transient, early-time dynamics and becomes negligible at sufficiently high temperatures since thermal fluctuations mask acceleration-induced effects. Throughout the evolution the measurement uncertainty and the retained quantum coherence remain approximately anti-correlated.The acceleration exerts a pronounced influence on the quantum features of the probe in ou cosmological backgrounds. The observer's acceleration governs the strength of thermal noise and degrades quantum coherence and thereby raises the measurement uncertainty.

 By contrast, an increase of the detector energy gap \(\omega\) tends to amplify the oscillatory amplitude and frequency, thereby making coherent oscillations more visible.This enhancement increases the sensibility of the probe and also significantly strengthens the information transfer toward it, which greatly improves its resilience against thermal noise.This observation underscores the utility of
the open-quantum systems approach, since the transient enhancement is only apparent
when the full detector dynamics are taken into account with the geometry of the cosmological background.When a temporal and thermal maximum exists, optimal temperature \(T_{\mathrm{opt}}\) and time estimation \(t_{\mathrm{opt}}\) can be achieved  before the probe reaches its stationary state.This point will be explicitly verified in the following subsection.

\subsection{Optimal temporal and thermal estimation}

The existence of an optimal interaction time and an optimal temperature is a key feature of quantum thermometry. At short times, the probe has not yet efficiently encoded thermal information, resulting in a low estimation precision. As the interaction time increases, both the quantum Fisher information and the quantum signal-to-noise ratio grow, reaching a maximum at a finite optimal time . A similar behavior is observed for the optimal temperature: high sensitivity is achieved in a finite temperature window, while at higher temperatures the system approaches thermal equilibrium and the sensitivity saturates to a constant value, becoming independent of the underlying geometry and environmental details.both the quantum Fisher information and the quantum signal-to-noise ratio grow, reaching a maximum at a finite optimal time \(t_{\mathrm{opt}}\) \(T_{\mathrm{opt}}\)

As stated earlier, the emergence of finite temporal and thermal peaks in the quantum Fisher information signals an optimal reactivity of the system. We analyze how the internal parameters of the AdS spacetime affect the optimal temperature and  time.

\begin{figure}
\includegraphics[width=1.0\linewidth]{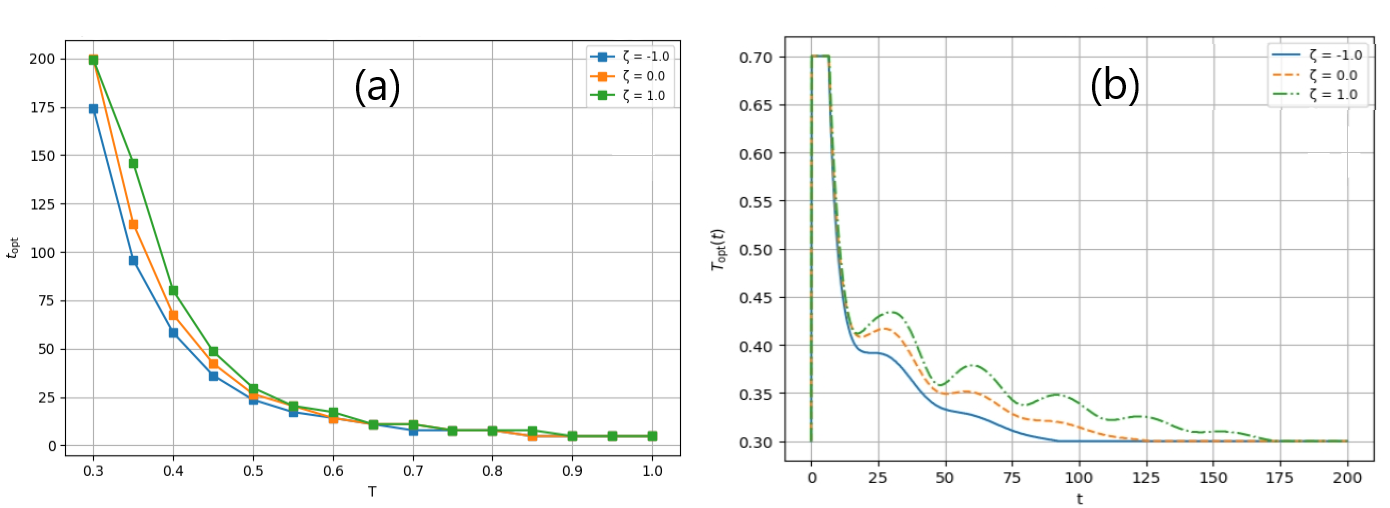}
\includegraphics[width=1.0\linewidth]{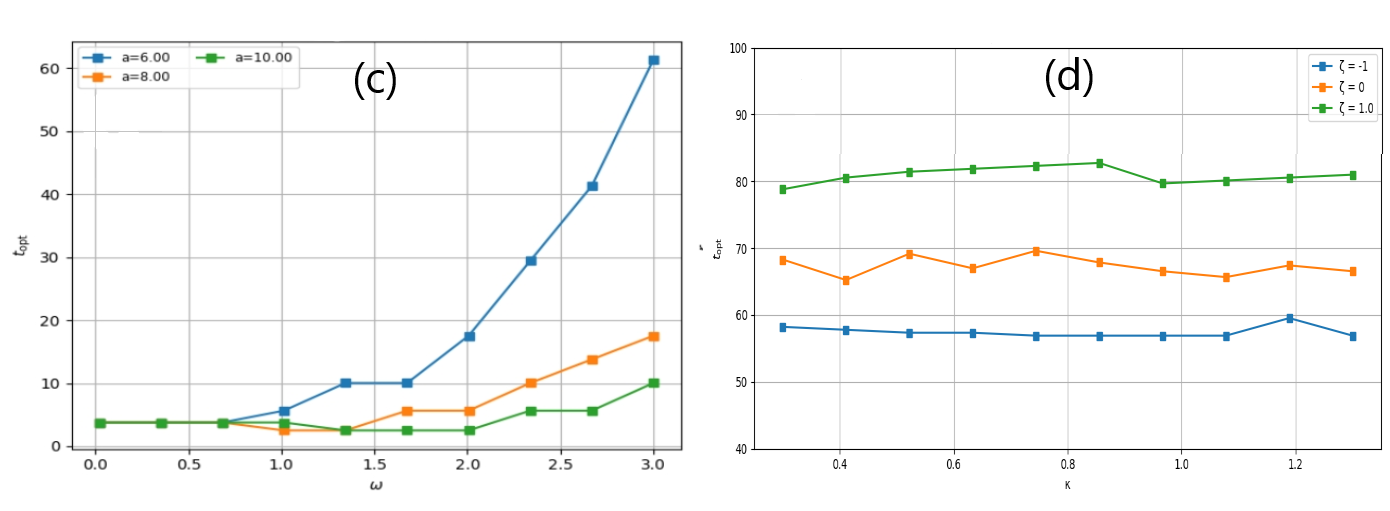} 
        \caption{ We analyze how the internal parameters , the energy gaps \(\omega\) ,  accelerations \(a\) and boundary conditions \(\zeta\) on the probe affect the optimal interaction time $t_{\mathrm{opt}}$ and the optimal temperature $T_{\mathrm{opt}}$, as well as their impact on the optimal quantum signal-to-noise ratio (QSNR) bound.}
    \label{fig1a}
\end{figure}

\begin{figure*}[t] 
        \includegraphics[width=5.9cm]{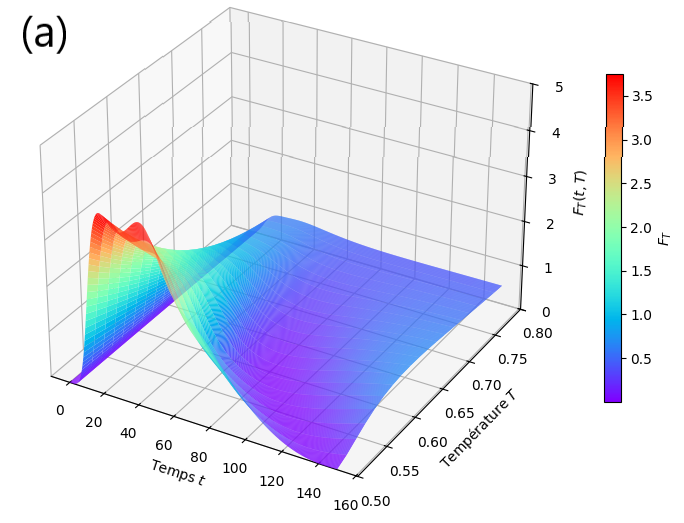}
        \includegraphics[width=5.9cm]{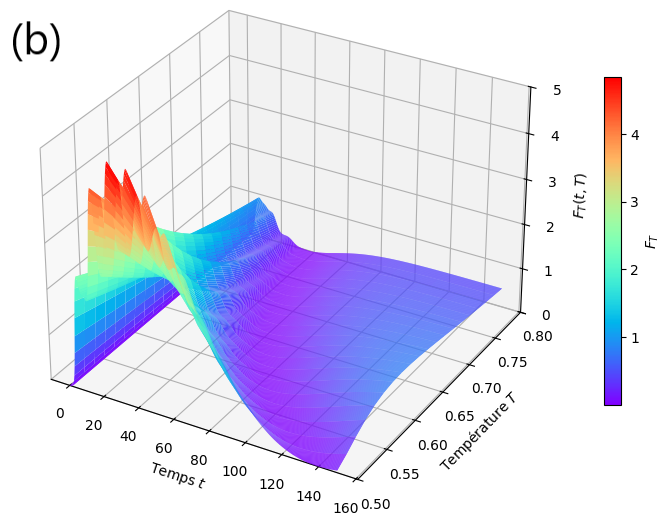}
        \includegraphics[width=5.9cm]{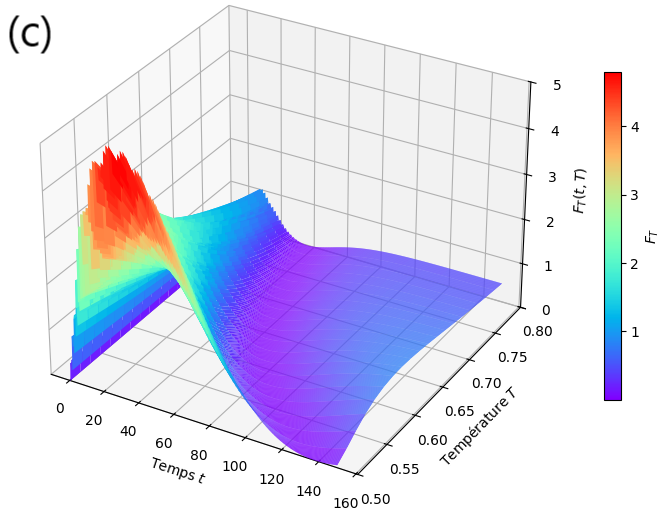}

      \caption{We show the QFI $H(T, t)$ as a function of both the reservoir temperature $T$ and the interaction time $t$ at the Dirichlet boundary condition (\(\zeta = 1\))for three illustrative types of structured environments:for three different coupling strengths \(\kappa\). Panels (a)–(c) correspond to \(\kappa = 0.1\), \(\kappa = 0.5\), and \(\kappa\ = 0.99\), respectively.}
    \label{figy}
\end{figure*}

The upper plots in fig \ref{fig1a}.a-b show the optimal interaction time \(t_{\mathrm{opt}}\)   as a function of the reservoir temperature and time interaction for different ADS boundary's. We remark different thermal regimes, where at low temperature the optimal interaction time (fig \ref{fig1a}-a) is comparatively long, whereas \(t_{\mathrm{opt}}\) increases as the \(\zeta\) rises , the short-time differences between curves for different boundary condition can be explained by the fact that larger \(\zeta\) tends to enhance the residual coherence and reduce the uncertainty during the transient dynamics.However,increasing the temperature progressively suppresses the influence of the boundary conditions. It is traced to the fact that  boundary-dependent contribution ,  contains a term proportional to
$
k\,\zeta\,\frac{1}{\pi T},$
that tends to zero at high \(T\), thereby erasing the \(\zeta\)-dependent correction and the residual coherence.the temperature \(T_{\mathrm{opt}}\) that maximizes the QFI for a given interaction time \(t\) in \ref{fig1a}-b , is particularly relevant for practical scenarios with a limited interaction window , since it identifies the temperature range that can be most accurately estimated within the available probing time.
which
summarize the parameter regimes where the probe achieves maximal sensitivity.We observe that the optimal temperatures are high at early times but then drop sharply as the system approaches thermalization. Close to thermal equilibrium the optimal temperature ceases to depend on the spacetime geometry and asymptotically converges to a constant value.

However, this attenuation proceeds at a slower rate for the Dirichlet boundary condition realized when \(\zeta=1\). The configuration  is more resilient to decoherence and thermalizes more slowly; consequently it provides enhanced protection to the UDW auxiliary detector against information loss .

We have analysed the influence of the boundary conditions of the \(\mathrm{AdS}_4\) spacetime on the estimation of both temporal and thermal peaks; in the sequel we fix \(\zeta=1\) and systematically investigate the roles of the probe acceleration \(a\), the detector energy gap \(\omega\), and the probe–Unruh–DeWitt auxiliary coupling \(\kappa\) to characterise how these internal parameters modulate the nature and rates of thermal exchanges in both thermalised and non-equilibrium regimes, to extract relevant time scales such as coherence times and an effective thermalisation that improve temporal and thermal peak estimation .

The results  in \cite{ref108} showed that the quantum uncertainty  measured by a comoving Unruh--DeWitt detector in de Anti-Sitter spacetime depends  on the ratio \(\omega/T\), where \(\omega\) denotes the detector energy gap and \(T\) the effective temperature The two parameters considered above also affect the detector's effective acceleration \(a\). In particular, an increase of the environmental temperature \(T\) modifies the detector's response and can be associated with a change in the apparent acceleration experienced by the probe. Indeed, a uniformly accelerated detector in \(\mathrm{AdS}_{4}\) perceives a nonvanishing temperature, so thermal and kinematic effects are intimately connected in this geometry 
, where increasing the acceleration is accompanied by an enhancement
in the effective temperature, which in turn leads to a faster thermalization and a smaller
optimal interaction time. Moreover, an increase of the probe energy gap \(\omega\) strengthens the probe–field interaction and thereby alters both the magnitude and the character of the coupling between the probe and the cosmological background , as a consequence, the optimal interaction time becomes shorter, while the thermal sensitivity of the probe is significantly reduced for high accelerations.
The coupling strength $\kappa$ between the probe and the auxiliary Unruh--DeWitt (UDW) detector produces only a slight change in the optimal interaction time, indicating that thermal and temporal peaks are not shifted but are mildly deformed .As will be demonstrated later , the rise of the coupling strength $\kappa$  increases  the probe’s non-Markovianity   , that enhances the probe's  sensitivity by amplifying oscillations and fine structure in the response curves (QFI,QSNR) , which consequently become less smooth .

In summary, our single-qubit thermometer operating in a structured, non-Markovian environment attains high precision as quantified by the QSNR, showing superior low-temperature sensitivity relative to previously studied protocols such as the dephasing-based thermometer \cite{39,40} . We note, however, that those alternative approaches may offer the practical advantage of shorter encoding times, whereas the present scheme trades somewhat longer interaction durations for improved accuracy at low temperatures.

\section{NON-MARKOVIAN QUANTUM THERMOMETRY}
\label{X6}
The composite environment is partitioned into two constituents: an ancilla subsystem and a Markovian reservoir, and this structure can induce non-Markovian dynamics on the probe. We obtain the joint state of the probe and ancilla by solving the master equation for the composite system exactly.
In this protocol, the probe encodes thermal information about the composite reservoir and thus serves as a quantum thermometer.The performance of temperature estimation is quantified by the quantum signal-to-noise ratio (QSNR) \cite{19}. In this section we investigate the impact of non-Markovian dynamics induced by the inter-probe–ancilla coupling \(\kappa\) on the quantum Fisher information (QFI), denoted \(\mathcal{F}_T\), and on the quantum signal-to-noise ratio (QSNR), denoted \(R_T\). Our analysis covers both the thermalized and the non-equilibrium regimes: we show how the coupling \(\kappa\) modulates the thermal sensitivity of the auxiliary system and how this modulation is subsequently transmitted to the probe.

\subsection{Non-Markovianity for a probe in the AdS space-time}

\begin{figure}
    \centering
\includegraphics[width=1.0\linewidth]{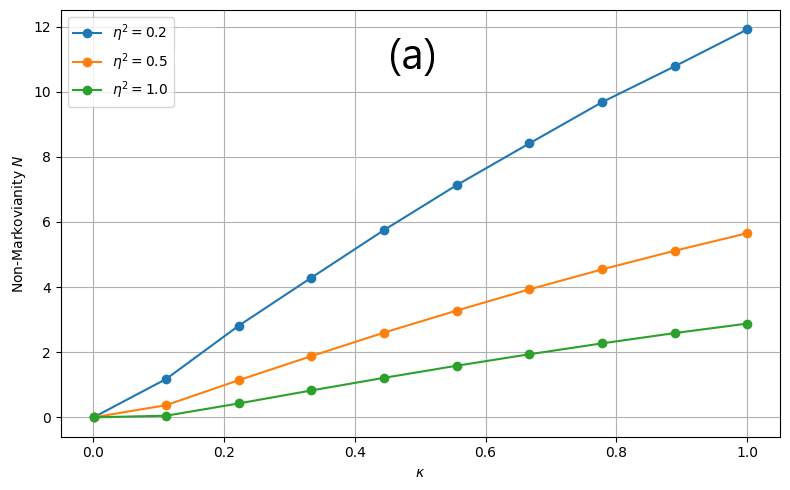}
\includegraphics[width=1.0\linewidth]{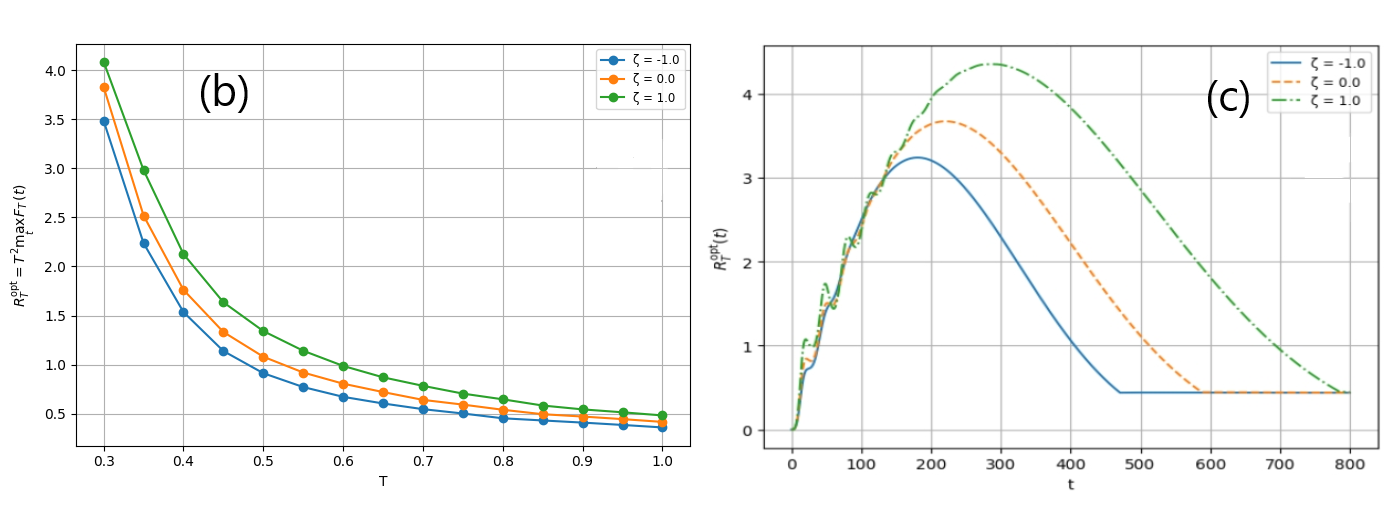} 
    \caption{Non-Markovianity \(\mathcal{N}\) as a function of the coupling strength \(\kappa\) for several values of the system--environment coupling \(\eta^2\). Specifically, \(\eta^2 = 1.0\), \(0.5\), and \(0.2\) correspond to the green , orange dashed, and blue solid curves, respectively.We analyze also the impact of the boundary conditions \(\zeta\) on the quantum signal-to-noise ratio (QSNR) boun. by analyzing the optimal interaction time $t_{\mathrm{opt}}$ at the optimal temperature $T_{\mathrm{opt}}$.}
    \label{figx}
\end{figure}

To quantify the presence of memory effects for the probe, we adopt an information-flow criterion similarly to \cite{14,15,16} . Where the trace distance quantifies how well the states can be distinguished: a monotonically decreasing $D(t)$ corresponds to a continuous loss of information from the system to the environment and hence Markovian behaviour \cite{aaa}, whereas any temporal increase of $D(t)$ signals information backflow from the environment to the system, which is the hallmark of non-Markovian dynamics \cite{bbb,ccc}.
For two-level probes the optimal initial pair maximizing the measure can be chosen as orthogonal pure states; a convenient choice is the pair $\ket{\pm}=\frac{\ket{0}\pm\ket{1}}{\sqrt{2}}$ , which has been widely used in previous studies.The fig-\ref{figx}-a presents the non-Markovianity measure \(\mathcal{N}\) plotted against the inter-environment coupling \(\kappa\) for several values of the system–environment coupling \(\eta\) that controls

the character
of the quantum field vacuum fluctuations and plays an important role in the dynamics of
the auxiliary UDW detector through the probe-field interaction.

For small values of \(\kappa\) we find \(\mathcal{N}=0\), indicating effectively Markovian dynamics in that regime. As \(\kappa\) increases the measure departs from zero, signalling the onset of memory effects.

 Interestingly, the magnitude of Markovianity diminishes as \(\eta\) grows, which implies that the composite environment attains a strongly non-Markovian character only for sufficiently large \(\kappa\) combined with relatively small \(\eta\).We next examine the role of non-Markovian effects in temperature estimation and
sensitivity .

To evaluate the thermal behavior of the probe and the effect of non-Markovianity on temperature estimation for different values of the interaction strength \(\kappa\),in  Fig-\ref{figy} the QFI is analyzed as a function of the temperature \(T\) and the interaction time \(t\) to identify the internal optimal conditions leading to improved temperature sensitivity.For early time cases a clear local maximum of \(H\) in time is visible at weak \(T\) and that an increase in the temperature 
T disrupts the thermal sensitivity transmitted by the UDW detector to the probe because increasing the temperature \(T\) raises the steady-state uncertainty \(U_{s}\) while suppressing the residual coherence \(C_{s}\), indicating that a finite optimal probing time and temperature exists , the results also provide a compelling evidence that larger values of the coupling \(\kappa\) yield a more pronounced 
 oscillatory structure of \(F(t,T)\)  becomes more prominent as \(\kappa\) increases , indicating a strong \(\kappa\)-dependence  . These findings demonstrated that non-Markovian dynamics , improve the precision of temperature estimation outside equilibrium. Nonetheless, in the long-time limit the QFI relaxes to a  value that depends only on the temperature of the composite environment and is effectively independent of \(\kappa\).The probe's
capacity to acquire information about the ambient temperature derives from its
sensitivity to thermalization: at low temperature decoherence processes are weak and it
therefore requires a longer time for temperature-dependent signatures to become
imprinted on the probe, whereas at high temperature decoherence proceeds rapidly
because, in this  regime thermal fluctuations dominate the decoherence
dynamics and the microscopic structure of the environment becomes effectively
irrelevant.

\begin{figure*}[t]

        \includegraphics[width=5.8cm]{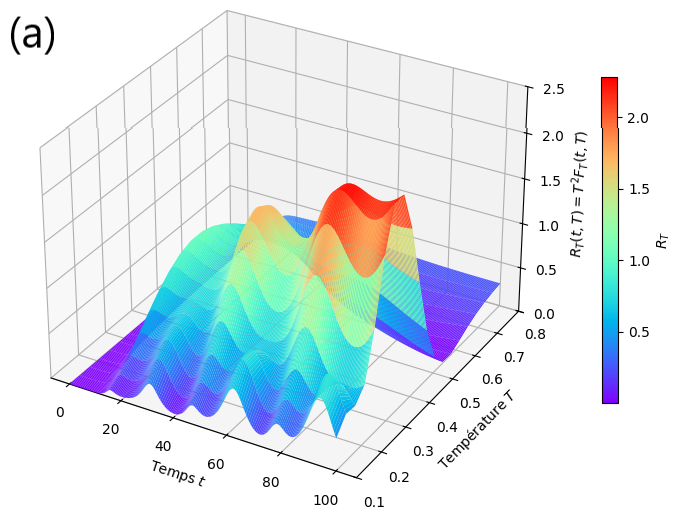}
        \includegraphics[width=5.8cm]{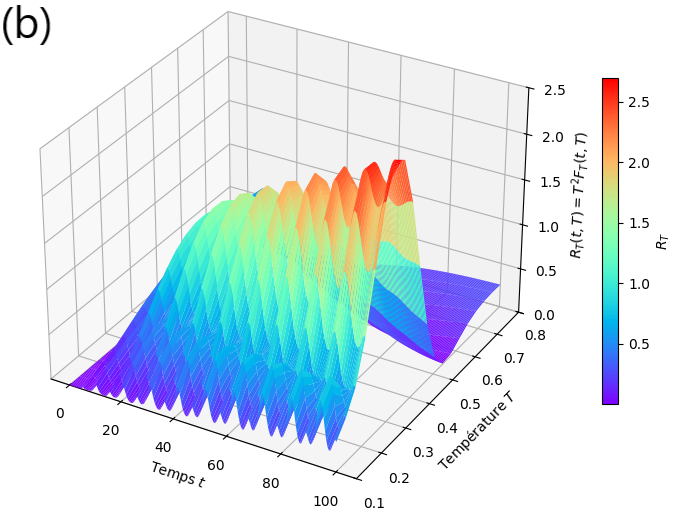}
        \includegraphics[width=5.8cm]{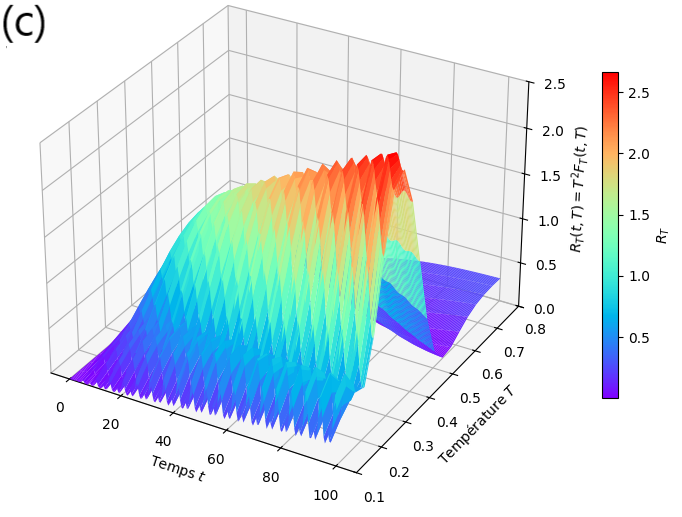}
      \caption{  The quantum signal-to-noise ratio (QSNR) is displayed as a function of the interaction time \(t\) and the temperature \(T\). at the Dirichlet boundary condition (\(\zeta = 1\)) for three different coupling strengths \(\kappa\). Panels (a)–(c) correspond to \(\kappa = 0.1\), \(\kappa = 0.5\), and \(\kappa= 0.99\).}
    \label{fig77}
\end{figure*}

\subsection{Optimal Bound}

In many sensing scenarios, the physical quantities of interest cannot be accessed directly, making indirect estimation strategies indispensable. Estimation theory provides a systematic framework to infer such quantities with high precision by analyzing measurement data associated with alternative observables.

We begin by reviewing the basic concepts of quantum thermometry. In our approach, the temperature—being a parameter rather than a quantum observable—is inferred through measurements performed on a quantum probe, as schematically illustrated in fig.~\ref{fig00}. The quantum Fisher information (QFI) plays a central role in this framework, as it quantifies the ultimate precision attainable in any thermometric protocol. According to the quantum Cramér--Rao bound, the variance of any unbiased temperature estimator $\hat{T}$ 
By invoking the quantum Cram\'er–Rao bound one obtains the fundamental inequality
where $H(T)$ denotes the quantum Fisher information (QFI) with respect to the parameter $T$. The quantity $Q_{T}$ is customarily referred to as the quantum signal-to-noise ratio (QSNR): a larger QSNR signals that the temperature $T$ can be estimated more effectively.
In what follows we shall first examine the effect of periodic boundary conditions on the quantum signal-to-noise ratio (QSNR) at the optimal time and temperature and then analyse the behaviour of the QSNR as a function of \(t\) and \(T\) for different probe–ancilla coupling strengths \(\kappa\), with the aim of elucidating how non-Markovian dynamics modify the uncertainty bound and the estimation performance of the detector.

Fig \ref{figx}-b addresses the overall temperature estimability: we plot the quantum signal-to-noise ratio (QSNR) evaluated at the optimal Temperature $T_{\mathrm{opt}}$, as a function of t. At early times  , the QSNR is essentially zero,indicating that thermal estimation is highly inefficient in this regime. As the temperature increases, the QSNR grows accordingly. In the intermediate temperature range, its behavior depends on the border parameter 
\(\zeta\); however, in the high\textendash temperature limit, the detailed properties of the environment no longer play a significant role, and the QSNR converges to a universal saturation value that is independent of the specific form of the environmental boundary $\zeta$ .

The fig \ref{figx}-c report the optimal interaction time \(t_{\mathrm{opt}}\)—as a function of temperature T. Different slopes in these curves mark distinct thermal regimes (notably low versus high temperature): \(t_{\mathrm{opt}}\) is relatively long at low temperatures and decreases as temperature rises. This trend follows from the probing mechanism: the probe encodes temperature information via decoherence, which is weak and slow at low \(T\) but accelerates at higher \(T\). In the high-temperature limit decoherence is dominated by thermal fluctuations and becomes insensitive to the environment's geometry. By contrast, at low temperature the environment's structure governs the decoherence dynamics and yields a pronounced \(\zeta\)-dependence of the QSNR particularly for Dirichlet's boundary . As shown above, the contribution of the parameter \(\zeta\) to the response function is inversely proportional to the temperature. In particular one may write, up to a model-dependent prefactor,
\[
\mathcal{R}_\zeta(T)\propto \frac{\zeta}{T},
\]
so that
\[
\lim_{T\to\infty}\mathcal{R}_\zeta(T)=0.
\]
Consequently, in the high-temperature limit the \(\zeta\)-dependent term disappears, which leads to a suppression of boundary  effects in the high thermalised regimes. As a result the system exhibits a qualitatively similar behaviour across different values of \(\zeta\) and an asymptotic convergence toward the same thermal response for all \(\zeta\).

The fig \ref{fig77} displays the time-dependent behavior of the quantum signal-to-noise ratio \(R(t,T)\) as a function of the interaction time \(t\) and the composite-environment temperature \(T\), for several values of the coupling strength \(\kappa\). The results indicate that stronger coupling markedly enhances the QSNR relative to weaker coupling, revealing a clear dependence of the signal quality on  \(\kappa\) under non-equilibrium conditions. Moreover, the oscillatory features of \(R(t,T)\)  become increasingly pronounced as  \(\kappa\) increases. This effect is particularly striking near \( \kappa\approx 0.99\) , these observations suggest that non-Markovian dynamics can be exploited to improve temperature-sensing performance away from equilibrium. Nevertheless, for any fixed \(\kappa\)the QSNR settles to the same asymptotic value at long times, which depends only on the environment temperature and the best temperature resolution in this scenario is achieved in the very low-temperature regime, as illustrated in Fig \ref{fig77}.

In our scheme the presence of non-equilibrium dynamics does not change the ultimate achievable accuracy of temperature estimation: the QSNR attains its maximum as the probe approaches its steady state. The connection between the probe qubit’s relaxation to steady state and the maximization of the QSNR has important implications for quantum thermometry: it underlines that probe stability is a key factor controlling both precision and sensitivity. As the probe equilibrates, the QSNR increases, making the probe more responsive to small temperature variations and improving measurement precision. These findings provide practical guidance for enhancing quantum thermometry protocols and establish a direct link between probe equilibration and temperature-measurement accuracy.

\section{Conclusion}

In this work, we have investigated single-qubit quantum thermometry, with particular attention to memory effects, demonstrating the probe's capability to estimate the temperature of a cosmological environment composed of a Markovian reservoir coupled to an ancillary system.  We have also examined the capability of Unruh--DeWitt detectors to probe both quantum uncertainty and quantum coherence in AdS spacetimes \cite{z10} by studying the detector response functions across several configurations. Remarkably, the asymptotic values  in our spacetime  depend only on the dimensionless ratio \(\omega/T\) (with \(\omega\) the detector energy gap and \(T\) the effective temperature): larger \(\omega/T\) corresponds to greater residual coherence and lower asymptotic uncertainty. Moreover, the results reveal a clear negative correlation between uncertainty and coherence.\par

Temperature plays a dominant role in degrading quantumness in both backgrounds. An increase in acceleration rises the temperatur \(T\) and amplifies thermal noise, which suppresses coherence and increases uncertainty \cite{z11}. For probes initially prepared in superposition states, both uncertainty and quantum correlations exhibit oscillatory dynamics at low temperatures; these oscillations are progressively damped as \(T\) increases, while enlarging the energy gap \(\omega\) tends to enhance both the amplitude and frequency of the oscillations.\par

The detector's acceleration significantly modifies the early-time dynamics of uncertainty and coherence, but it does not affect their long-time stationary values. The influence of acceleration weakens with increasing temperature and becomes negligible in the high-\(T\) regime. Analogous transient modifications originate from different boundary prescriptions in AdS \cite{z12}, although the late-time values remain essentially unchanged, wich typically display greater sensitivity and thermalize more rapidly under identical thermal conditions, thereby reducing the impact of ancillary factors (for instance, distinct boundary conditions) on the measured uncertainty and coherence. The AdS geometries the detector ultimately thermalizes with the surrounding field under prolonged interaction \cite{z13}, yielding stationary values for the uncertainty and for the quantum coherence.We find also that non-Markovianity is strongly affected by the UDW's nonequilibrium dynamics: it increases with the inter-environment coupling \(\kappa\) and produces rapid, oscillatory enhancements of the QSNR, in contrast to the slower growth characteristic of the effectively Markovian regime. Nevertheless, the asymptotic QSNR in the long-time limit depends solely on the temperature of the composite environment. The peak QSNR as a function of temperature exhibits a single maximum located in the low-temperature region, suggesting that the proposed protocol is particularly well-suited for the design of low-temperature quantum sensors. We also analysed the situation where the composite environment features distinct temperatures and studied how the Markovian reservoir affects the ancilla's temperature estimation. For each fixed ancilla temperature the QSNR over time displays a clear maximum, and we highlighted how non-Markovian effects can both enhance the estimation efficiency and shorten the optimal interaction time. A stable (higher) temperature in the Markovian reservoir tends to increase the peak QSNR for ancilla-temperature estimation relative to the low-temperature case. These observations shed light on the interplay between geometry, thermal noise and detector dynamics, and suggest that UDW-type probes may provide useful diagnostic tools for quantum-information tasks formulated in curved-spacetime settings.

\section*{ACKNOWLEDGMENTS}

A.H acknowledges the financial support of the National Center for Scientific and Technical Research (CNRST) through the "PhD-Associate Scholarship-PASS" program. The authors acknowledge the LPHE-MS, FSR for the technical support.\par 

\textbf{Declaration of competing interest:}\par 
The authors declare that they have no known competing financial interests or personal relationships that could have appeared to influence the work reported in this paper.\par

\textbf{Data availability:}\par 
No data was used for the research described in the article.

\appendix

\section{Master equation}
\label{X7}
The unitary evolution of the global state $\rho_{\rm tot}$ is governed by the Liouville--von Neumann equation
\begin{equation}
\frac{\partial \rho_{\rm tot}}{\partial \tau} = -i\,[H,\rho_{\rm tot}],
\end{equation}
with initial condition $\rho_{\rm tot}(0)=\rho_D(0)\otimes |0\rangle\langle 0|$, where $\rho_D(0)$ is the detector initial state and $|0\rangle$ denotes the conformal vacuum of the scalar field $\varphi(x)$. Tracing out the field degrees of freedom yields the reduced state of the detector, $\rho_D(\tau)=\operatorname{Tr}_{\varphi}[\rho_{\rm tot}(\tau)]$.

In the weak-coupling, Markovian limit the reduced dynamics of the detector can be written in Kossakowski--Lindblad form:
\begin{equation}
\frac{\partial \rho_D(\tau)}{\partial \tau} = -i\,[H_{\rm eff},\rho_D(\tau)] + \mathcal{L}[\rho_D(\tau)],
\label{eq:lindblad}
\end{equation}
where the dissipator $\mathcal{L}$ takes the standard structure
\begin{equation}
\mathcal{L}[\rho] = \frac{1}{2}\sum_{i,j=1}^{3} C_{ij}\bigl(2\sigma_j\,\rho\,\sigma_i - \sigma_i\sigma_j\,\rho - \rho\,\sigma_i\sigma_j\bigr),
\end{equation}
and $\{\sigma_i\}_{i=1}^3$ are the Pauli matrices acting on the detector.

The Kossakowski matrix $C_{ij}$ can be written explicitly (in the chosen basis) as
\begin{equation}
C_{ij}=\begin{pmatrix}
A'-iB' & 0  & 0 \\[4pt]
iB' & A' & 0 \\[4pt]
0 & 0 & A'+C
\end{pmatrix},
\end{equation}
where the scalar coefficients are expressed in terms of the noise kernel $G(\omega)$ as
\begin{align}
A' &= \tfrac{1}{2}\bigl[G(\Omega)+G(-\Omega)\bigr], \\
B' &= \tfrac{1}{2}\bigl[G(\Omega)-G(-\Omega)\bigr], \\
C  &= G(0)-A'.
\end{align}

The field also renormalizes the detector frequency: the effective Hamiltonian of the detector reads
\begin{equation}
H_{\rm eff}=\tfrac{1}{2}\,\tilde{\Omega}\,\sigma_z,
\end{equation}
with the renormalized gap
\begin{equation}
\tilde{\Omega}=\Omega+\frac{1}{i}\bigl[K(-\Omega)-K(\Omega)\bigr],
\qquad
K(\Omega)=\frac{1}{\pi}\,\mathcal{P}\!\int_{-\infty}^{\infty} d\omega\;\frac{G(\omega)}{\omega-\Omega},
\end{equation}
where $\mathcal{P}$ denotes the Cauchy principal value and $G(\omega)$ is the Fourier transform of the Wightman function.

Assuming the detector is initially prepared in the pure state
\begin{equation}
|\psi\rangle=\cos\!\Bigl(\tfrac{\theta}{2}\Bigr)\,|0\rangle+\sin\!\Bigl(\tfrac{\theta}{2}\Bigr)\,|1\rangle,
\end{equation}
the master equation \eqref{eq:lindblad} can be solved analytically. The detector state at time $\tau$ admits the Bloch representation
\begin{equation}
\rho_D(\tau)=\tfrac{1}{2}\bigl(\mathbb{I}+\mathbf{n}(\tau)\!\cdot\!\boldsymbol{\sigma}\bigr),
\end{equation}
with Bloch vector components
\begin{align}
n_1(\tau) &= e^{-A'\tau/2}\,\sin\theta\cos(\tilde{\Omega}\tau),\\
n_2(\tau) &= e^{-A'\tau/2}\,\sin\theta\sin(\tilde{\Omega}\tau),\\
n_3(\tau) &= e^{-A'\tau}\cos\theta - R\bigl(1-e^{-A'\tau}\bigr),
\end{align}
where we have introduced the ratio $R=B'/A'$.

\end{document}